\documentclass[12pt]{article}
\pdfoutput=1

\usepackage{tikz-cd}
\usepackage{amssymb,amsmath}
\usepackage{graphicx}
\usepackage{epsfig}
\usepackage{epstopdf}
\usepackage{caption}
\usepackage{subcaption}
\usepackage{latexsym}
\usepackage{psfrag}
\usepackage{booktabs}
\usepackage{bbm}
\usepackage{mathrsfs}

\usepackage[toc]{appendix}

\usepackage{color}
\usepackage{datetime}
\usepackage[
      colorlinks=false,
      linkcolor=darkblue,  
      urlcolor=blue,    
      filecolor=blue,     
      citecolor=red,
linktocpage=true,
      pdfstartview=FitV,
      bookmarksopen=true    
      ]{hyperref}

\usepackage[capitalize]{cleveref}
\crefname{equation}{}{}
\Crefname{equation}{}{}

\ifpdf
\fi

\numberwithin{equation}{section}  

\def\coeff#1#2{\relax{\textstyle {#1 \over #2}}\displaystyle}

\def\ZZ{\mathbb{Z}}

\def\cF{{\cal F}}

\def\cO{{\cal O}}

\def\cQ{{\cal Q}}
\def\cR{{\cal R}}

\def\Neql#1{{\cal N}\!=\!{#1}}

\def\RR{\mathbb{R}}

\def\HH{\mathbb{H}}

\definecolor{cardinal}{rgb}{0.6,0,0}
\definecolor{darkgreen}{rgb}{0,0.5,0}
\definecolor{golden}{rgb}{0.92, 0.7, 0}
\definecolor{midnight}{rgb}{0, 0, 0.5}
\definecolor{darkblue}{rgb}{0.2, 0, 0.8}

\newcommand{\dd}{\mathrm{d}}					
\DeclareMathOperator*{\hodge}{\star}				
\newcommand{\vol}{\mathrm{vol}}				

\DeclareMathOperator*{\diag}{\mathrm{diag}}		
\DeclareMathOperator{\laplace}{\Delta}			

\newcommand{\abs}[1]{\lvert {#1} \rvert}			
\DeclareMathOperator{\sign}{sign}				

\usepackage{forloop}
\newcounter{ct}

\newcommand{\eucl}[1]{(%
	\ifthenelse{#1 > 0}{%
		\mathord{+} \forloop[-1]{ct}{#1}{\value{ct} > 1}{\, \mathord{+}}%
	}{}%
)}

\newcommand{\sig}[2]{(%
	\ifthenelse{#2 > 0}{%
		\mathord{-}  \forloop[-1]{ct}{#2}{\value{ct} > 1}{\, \mathord{-}}%
		\ifthenelse{#1 > 0}{\,}{}%
	}{}%
	\ifthenelse{#1 > 0}{%
		\mathord{+} \forloop[-1]{ct}{#1}{\value{ct} > 1}{\, \mathord{+}}%
	}{}%
)}

\newcommand{\Om}[1]{\Omega^{(#1)}_-}
\newcommand{\Op}[1]{\Omega^{(#1)}_+}
\newcommand{\Opm}[1]{\Omega^{(#1)}_{\pm}}

\newcommand{\om}[1]{\omega^{(#1)}_-}

\newcommand{\Th}[1]{\Theta^{(#1)}}

\newcommand{\Flux}[1]{\Pi^{(#1)}}
\newcommand{\hFlux}[1]{\widehat{\Pi}^{(#1)}}

\newcommand{\flux}[1]{\Pi^{(#1)}}
\newcommand{\hflux}[1]{\widehat{\Pi}^{(#1)}}

\allowdisplaybreaks[1]

\begin{document}  

\begin{titlepage}
 
\begin{center}
\bigskip
\bigskip

{\Large \bf Non-Supersymmetric, Multi-Center Solutions with Topological Flux}

\medskip
\bigskip
\bigskip
{\bf Benjamin E. Niehoff} \\
\bigskip
Department of Physics and Astronomy \\
University of Southern California \\
Los Angeles, CA 90089, USA \\
\bigskip
\bigskip
{\rm bniehoff@usc.edu} \\
\bigskip
\bigskip
\end{center}

\begin{abstract}
\noindent We find an infinite class of non-supersymmetric multi-center solutions to the STU model in five-dimensional ungauged supergravity coupled to two vector multiplets.  The solutions are obtained by solving a system of linear equations on a class of Ricci-scalar-flat K\"ahler manifolds studied by LeBrun.  After imposing an additional $U(1)$ isometry in the base, we solve the axisymmetric $SU(\infty)$ Toda equation and obtain explicit supergravity solutions containing arbitrary numbers of 2-cycles with cohomological fluxes of all three flavors.  This improves upon a previous result where only two of the three fluxes were topologically non-trivial.  Imposing regularity and absence of closed timelike curves, we obtain ``bubble equations" highly reminiscent of those known in the supersymmetric case.  Thus we extend much of the analysis done for BPS bubbling solutions to this new family of non-supersymmetric bubbling solutions.
\end{abstract}

\end{titlepage}


\tableofcontents

\section{Introduction}

In the past few years, there have been many exciting developments in the program of finding black hole microstate geometries.  These are solitonic solutions to supergravity theories which have the same asymptotic behavior as a given black hole (or black ring), including mass, charge, and angular momentum, and yet in the bulk remain totally smooth and free of horizons.  Instead, the pathological parts of the would-be black hole are resolved by a collection of smooth, topological bubbles, threaded by cohomological fluxes which hold the whole thing up against gravitational collapse.  It is conjectured that such geometries may provide the ``hair" necessary to store the entropy of the black hole (or black ring) \cite{Bekenstein:1973ur, Hawking:1974rv}, and can be interpreted as supergravity approximations to the stringy states (or ``fuzzballs") thought to resolve the information paradox \cite{Mathur:2005zp}.  Beyond specifically finding smooth microstate geometries, this program is of general interest for providing numerous examples of stationary supergravity solutions containing arbitrary collections of charged, rotating black holes and rings balanced by their mutual electromagnetic interactions.

Of central importance to this program is the discovery that the BPS equations for 5-dimensional, $\Neql 2$ supergravity coupled to vector multiplets can be cast as a \emph{linear system} \cite{Bena:2004de}.  From this came a whole body of work on BPS solutions, extending previously-known families of solutions and uncovering new ones; especially leading to the construction of the ``bubbling microstate geometries", or solitons made of pure topological bubbles and fluxes \cite{Bena:2005va, Berglund:2005vb, Bena:2007kg, Gibbons:2013tqa}.  These solutions are constructed with a time fiber over a hyper-K\"ahler Gibbons-Hawking (GH) base \cite{Gibbons:1979zt}, which contains topologically non-trivial 2-cycles supported by harmonic fluxes.  In particular, one finds a set of ``bubble equations", which arise from demanding the absence of closed timelike curves.  The bubble equations relate the cohomological fluxes to the sizes of the homological bubbles to which they are linked; thus, the bubbles are literally held open by the fluxes.

More recently, there have been several attempts to get away from BPS.  A few isolated examples exist \cite{Jejjala:2005yu, Gimon:2007ps, Giusto:2007tt, Bena:2009qv} of truly non-BPS, non-extremal smooth geometries, but no infinite familes are yet known (which are necessary for entropy counting).  However, in the non-BPS \emph{extremal} case, there are linear systems which can be solved to obtain infinite families of solutions.  One such family are the so-called ``almost BPS" solutions \cite{Goldstein:2008fq, Bena:2009ev, Bena:2009en}, where supersymmetry is broken by inverting the orientation of the Gibbons-Hawking base relative to the fluxes.  These solutions have been shown to exhibit a rich variety of phenomena not seen in the BPS case \cite{Vasilakis:2011ki, Bena:2013gma, Chowdhury:2013ema}.

A further avenue of attack was revealed with the ``floating brane" ansatz in 5 dimensions, which dispenses with supersymmetry, but still imposes a generic balance between gravitational and electromagnetic forces.  It was found that this leads to yet another \emph{linear system} of equations, this time on a Euclidean-signature Einstein-Maxwell base \cite{Bena:2009fi}.  A few solutions are known based on various Euclidean-Einstein-Maxwell geometries analytically continued from classical GR ones \cite{Bobev:2009kn}, as well as an infinite family given in \cite{Bena:2009fi} based on the Israel-Wilson metric.

In a pair of recent papers \cite{Bobev:2011kk, Bobev:2012af}, the author and collaborators have presented an infinite family of ``floating brane" non-BPS solutions based on a family of K\"ahler Einstein-Maxwell metrics studied by LeBrun \cite{LeBrun:1991, LeBrun:2008kh}.  These metrics are determined by two functions which solve the $SU(\infty)$ Toda equation and its linearization.  By choosing an extremely simple solution to the Toda equation, one obtains the subclass of LeBrun-Burns metrics, which are K\"ahler analogues to Gibbons-Hawking metrics with a hyperbolic base instead of flat $\RR^3$.  On the LeBrun-Burns base, the floating brane equations are solvable and one obtains an infinite family of solutions.

These solutions were shown to have a few desirable properties.  The LeBrun-Burns metrics have the structure of a $U(1)$ fiber over $\HH^3$.  In much the same way as Gibbons-Hawking metrics, this $U(1)$ fiber pinches off at controlled points, which allows one to construct solutions with several ``bubbles" threaded with cohomological fluxes.  It was also shown that with appropriate choices of parameters, the solutions could be made regular and free of CTC's.

However, these solutions also had a few shortcomings.  The Maxwell field of the LeBrun-Burns metrics is topologically trivial.  Hence, while one can use the $U(1)$ fiber to form 2-cycles, only two of the three fluxes thread those 2-cycles.  The resulting ``bubble equations" turn out to be independent of the sizes of the bubbles, and thus the interplay between bubbles and fluxes is gone.  Furthermore, the solution is very degenerate, because it effectively has only two types of dipole charges.  As a result, the regularity conditions actually demand that most of the parameters be set to zero.  Finally, the solutions are not asymptotically flat; in fact, it was shown that the floating brane equations on a K\"ahler base have no asymptotically-flat solutions in general \cite{Bobev:2011kk}.  This last shortcoming should not be all too great a concern.  One does obtain solutions whose asymptotics are like the near-horizon limit of a BMPV black hole \cite{Breckenridge:1996is}.  So it is not too far a stretch to say that these are BMPV microstate geometries, and probably the asymptotic region can be restored by relaxing the assumptions of the floating brane ansatz.

Yet another linear system of equations was discovered by re-organizing the BPS equations in the 6-dimensional IIB frame \cite{Gutowski:2003rg, Cariglia:2004kk, Bena:2011dd}, which makes a curious connection to the 5-dimensional story: the 5-dimensional non-BPS, floating brane equations on a K\"ahler base are \emph{identical} to the 6-dimensional BPS equations where all functions are made independent of the 6th coordinate \cite{Bobev:2012af}.  Therefore the exact same family of solutions plays two roles, both supersymmetric and non-supersymmetric.  The apparent discrepancy is explained in the KK reduction from 6 to 5 dimensions: the Killing spinor in 6 dimensions can be charged under the $U(1)$ on which the reduction occurs, which causes it to vanish in 5 dimensions.  This is reminiscent of the Scherk-Schwarz mechanism \cite{Scherk:1979zr, Scherk:1978ta}, or also ``supersymmetry without supersymmetry" \cite{Duff:1997qz}.

In this paper, we improve upon the results of \cite{Bobev:2011kk} and overcome its major issues.  Despite the 5d-6d link mentioned, we work strictly in the 5-dimensional frame, as it is the simpler of the two.  This paper is organized as follows:   In Section 2, we briefly describe the 5d $\Neql 2$ theory, the floating brane ansatz, and the equations that result.  In Section 3, we describe the basic features of LeBrun metrics in general, and the system that results from putting the floating brane equations on the LeBrun base.  We show how the system is solved generically.  In Section 4, we solve the $SU(\infty)$ Toda equation explicitly under the assumption of an additional $U(1)$ isometry.  We determine the boundary conditions needed for the solutions we wish to build, and we analyze the resulting base manifold in detail to explore its geometric and topological properties.  In Section 5, we solve the floating brane equations on this base manifold explicitly, thus giving the full supergravity solution.  We determine the conditions needed to make solutions regular in 5 dimensions.  We derive the no-CTC conditions, or ``bubble equations" and analyze them.  Finally, we give an explicit, solved example of a 3-center solution.  In Section 6, we present our conclusions.

\section{Non-BPS solutions from floating branes}
\label{floating branes}

It is simplest to present our solutions in the context of $\Neql 2$ ungauged supergravity in 5 dimensions coupled to two vector multiplets (thus having three $U(1)$ gauge fields).  One can also see this theory as a truncation of eleven-dimensional supergravity on $T^6$.  The 5-dimensional action is
\begin{equation}
S = \frac{1}{2 \kappa_5} \int \bigg( \hodge_5 \cR - Q_{IJ} \, \dd X^I \wedge \hodge_5 \dd X^J - Q_{IJ} \, F^I \wedge \hodge_5 F^J - \frac16 \, C_{IJK} \, F^I \wedge F^J \wedge A^K \bigg),
\label{5d action}
\end{equation}
where $\cR$ is the Ricci scalar, $X^I, \; I \in \{ 1,2,3 \}$ are scalar fields, $F^I \equiv dA^I$ are three Maxwell fields, and the kinetic terms are coupled via the matrix
\begin{equation}
Q_{IJ} \equiv \frac12 \diag \big( (X^1)^{-2}, (X^2)^{-2}, (X^3)^{-2} \big).
\end{equation}
The scalar fields are subject to the constraint $X^1 X^2 X^3 = 1$, which we parametrize in terms of a new set of scalars $Z_I$ as
\begin{equation} \label{scalars}
X^1 = \bigg( \frac{Z_2 \, Z_3}{Z_1^2} \bigg)^{1/3}, \quad X^2 = \bigg( \frac{Z_1 \, Z_3}{Z_2^2} \bigg)^{1/3}, \quad X^3 = \bigg( \frac{Z_1 \, Z_2}{Z_3^2} \bigg)^{1/3}.
\end{equation}
These new scalars $Z_I$ are very convenient in the ans\"atze to follow.

We begin with the usual 5d metric ansatz,
\begin{equation} \label{5d metric}
\dd s_5^2 = - Z^{-2} \, (\dd t + k)^2 + Z \, \dd s_4^2, \qquad Z \equiv (Z_1 Z_2 Z_3)^{1/3},
\end{equation}
with 4d base manifold $\dd s_4^2$.  Following \cite{Bena:2009fi}, the Maxwell fields are then given by the ``floating brane" ansatz,
\begin{equation}
A^I \equiv - Z_I^{-1} \, (\dd t + k) + B^I,
\end{equation}
and it is convenient to introduce the magnetic 2-forms given by
\begin{equation}
\Th I \equiv \dd B^I.
\end{equation}
For completeness, we also give the embedding into 11-dimensional supergravity.  The 11-dimensional metric and 3-form potential are given by
\begin{align}
\begin{split}
\dd s_{11}^2 &= \dd s_5^2 + \bigg( \frac{Z_2 \, Z_3}{Z_1^2} \bigg)^{1/3} (\dd y_1^2 + \dd y_2^2) +  \bigg( \frac{Z_1 \, Z_3}{Z_2^2} \bigg)^{1/3} (\dd y_3^2 + \dd y_4^2) \\
& \qquad \qquad \qquad + \bigg( \frac{Z_1 \, Z_2}{Z_3^2} \bigg)^{1/3} (\dd y_5^2 + \dd y_6^2), \label{11d metric}
\end{split} \\
C^{(3)} &= A^1 \wedge \dd y_1 \wedge \dd y_2 + A^2 \wedge \dd y_3 \wedge \dd y_4 + A^3 \wedge \dd y_5 \wedge \dd y_6, \label{3form C}
\end{align}
where we see that the three scalars $X^I$ \eqref{scalars} come from the sizes of three $T^2$'s inside the $T^6$ spanned by the coordinates $y_i$.  In particular, for the $T^6$ to remain compact, the $Z_I$ must be everywhere finite and nonzero; or, if any of the $Z_I \to 0$ or $Z_I \to \infty$, they must all do so with the same behavior.

Returning to the 5-dimensional theory, as was shown in \cite{Bena:2009fi}, we then need a 4-dimensional base manifold that solves the Euclidean-signature Einstein-Maxwell equations,
\begin{equation}
{R}_{\mu\nu}^{(4)} = \coeff{1}{2}\, \big( \cF_{\mu\rho} {\cF_{\nu}}^{\rho} -  \coeff{1}{4}\, g_{\mu\nu} \cF_{\rho\sigma} \cF^{\rho\sigma} \big),
\end{equation}
where $\cF$ is a Maxwell 2-form determined by the base geometry, and unrelated to the $F^I$.  We decompose $\cF$ as
\begin{equation}
\cF \equiv \Th 3 - \om 3, \label{cF decomp}
\end{equation}
where $\Th 3$ is self-dual, and $\om 3$ is anti-self-dual.   The Maxwell equations $\dd \cF = \dd \hodge_4 \cF =0$ imply that $\Theta^{(3)}$ and $\omega^{(3)}_{-}$ are harmonic.  As the notation implies, this defines the magnetic 2-form field strength $\Theta^{(3)}$.

The equations of motion of \eqref{5d action} then reduce to the linear system \cite{Bena:2009fi}:
\begin{align}
\dd \hodge_4 \dd Z_1 &= \Th 2 \wedge \Th 3, & \Th 2 - \hodge_4 \Th 2 &= 2 \, Z_1 \, \om 3, \label{Z1 Th2 eqn} \\
\dd \hodge_4 \dd Z_2 &= \Th 1 \wedge \Th 3, & \Th 1 - \hodge_4 \Th 1 &= 2 \, Z_2 \, \om 3, \label{Z2 Th1 eqn}
\end{align}
and
\begin{align}
\dd \hodge_4 \dd Z_3 &= \Th 1 \wedge \Th 2 - \om 3 \wedge ( \dd k - \hodge_4 \dd k), \label{Z3 eqn} \\
\dd k + \hodge_4 \dd k &= \frac12 \sum_I Z_I \, ( \Th I + \hodge_4 \Th I ). \label{k eqn}
\end{align}
We solve the equations of motion by the following steps:  First, find a Euclidean-Einstein-Maxwell base.  The Maxwell 2-form defines the 2-forms $\Th 3$ and $\om 3$ via \eqref{cF decomp}.  We then solve the first layer of coupled linear equations \eqref{Z1 Th2 eqn} and \eqref{Z2 Th1 eqn} for $Z_1, Z_2, \Th 1,$ and $\Th 2$.  These enter as sources in the second layer of coupled linear equations \eqref{Z3 eqn} and \eqref{k eqn}, which we solve finally for $Z_3$ and $k$.  Next we follow \cite{Bobev:2011kk, Bobev:2012af} and implement this solution for the LeBrun metrics.

\section{LeBrun metrics}

The LeBrun family of metrics \cite{LeBrun:1991} is given by
\begin{equation}
g \equiv \frac{1}{w} (\dd \tau + A)^2 + w e^u \, (\dd x^2 + \dd y^2) + w \, \dd z^2, \label{lebrun}
\end{equation}
where $\tau$ is periodic with period $4 \pi$.  The functions $u$ and $w$ are independent of $\tau$ and solve the $SU(\infty)$ Toda equation and its linearization, respectively:
\begin{align}
u_{xx} + u_{yy} + (e^u)_{zz} &= 0, \label{toda} \\
w_{xx} + w_{yy} + (e^u w)_{zz} &= 0, \label{lintoda}
\end{align}
and the 1-form $A$ satisfies
\begin{equation}
\dd A = w_x \, \dd y \wedge \dd z + w_y \, \dd z \wedge \dd x + (e^u w)_z \, \dd x \wedge \dd y. \label{A eqn}
\end{equation}
Under the conditions \eqref{lintoda} and \eqref{A eqn}, the metric \eqref{lebrun} is K\"ahler, with K\"ahler form
\begin{equation}
J \equiv (\dd \tau + A) \wedge \dd z - e^u w \, \dd x \wedge \dd y.
\end{equation}
The condition \eqref{toda} further implies that the Ricci scalar vanishes \cite{LeBrun:1991}.

We choose to introduce the frames,
\begin{equation}
e^1 = w^{-1/2} \, (\dd \tau + A), \qquad e^2 = e^{u/2} w^{1/2} \, \dd x, \qquad e^3 = e^{u/2} w^{1/2} \, \dd y, \qquad e^4 = w^{1/2} \, \dd z,
\end{equation}
with orientation
\begin{equation}
\vol_4 \equiv e^1 \wedge e^2 \wedge e^3 \wedge e^4 = e^u w \, \dd \tau \wedge \dd x \wedge \dd y \wedge \dd z,
\end{equation}
such that $J$ is anti-self-dual.  It will also be helpful to define the (anti)-self-dual 2-forms
\begin{align}
\Opm{1} &= e^{-u/2} \, \big( e^1 \wedge e^2 \pm e^3 \wedge e^4 \big) &&= (\dd \tau + A) \wedge \dd x \pm w \, \dd y \wedge \dd z, \\
\Opm{2} &=  e^{-u/2} \, \big( e^1 \wedge e^3 \pm e^4 \wedge e^1 \big) &&= (\dd \tau + A) \wedge \dd y \pm w \, \dd z \wedge \dd x, \\
\Opm{3} &= e^1 \wedge e^4 \pm e^2 \wedge e^3 &&= (\dd \tau + A) \wedge \dd z \pm w e^u \, \dd x \wedge \dd y,
\end{align}
such that $J = \Om 3$.

\subsection{Topological structure}
\label{lebrun topo}

The LeBrun metrics \eqref{lebrun} have the structure of a $U(1)$ fiber over a 3-dimensional base with metric
\begin{equation}
h = e^u (\dd x^2 + \dd y^2) + \dd z^2,
\end{equation}
which in turn can be thought of as a Riemann surface fibered over a line.  If $e^u$ is everywhere finite and non-singular, then the $(x,y,z)$ coordinates can be extended to a topological $\RR^3$.  In this case, the topology of the 4-manifold can be analyzed in terms of the $U(1)$ fiber parametrized by $\tau$, much like the topology of Gibbons-Hawking manifolds \cite{Gibbons:1979zt}.

\begin{figure}
\centering
\includegraphics[height=3cm]{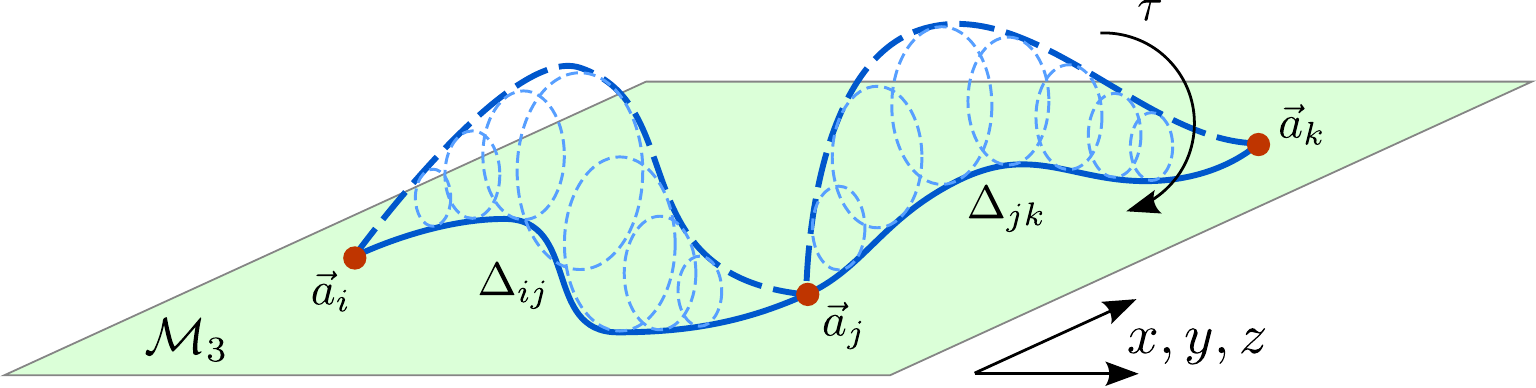}
    \caption{\it \small Homological 2-cycles in the LeBrun metric.  The $\tau$ fiber pinches off at the points $\vec a_i$.  Sweeping the fiber along a path between any two points forms a homological 2-sphere.  Two intersecting 2-cycles are shown.}
\label{cycles}
\end{figure}

The function $w$ solves a second-order Laplace-like equation, whose solutions are characterized by a number of points we will call ``Gibbons-Hawking points" or ``geometric charges", where locally (provided that $e^u$ is smooth),
\begin{equation}
w \sim \frac{1}{r},
\end{equation}
for some local radial distance $r$.  At these points the $\tau$ fiber pinches off to zero size.  Hence, if one takes any curve in the 3-dimensional base $h$ that joins two geometric charges, the surface described by the $\tau$ fiber over this curve is a homological 2-sphere, as in Fig. \ref{cycles}.

If $e^u$ is not smooth, it is still possible that $g$ is smooth.  One possibility is that $z$ is a radial coordinate, and $e^u (\dd x^2 + \dd y^2)$ describes a sphere (or perhaps a quotient of a sphere).  Another possibility is that $e^u (\dd x^2 + \dd y^2)$ is a higher-genus Riemann surface, in which case one can have topological cycles that do not involve the $\tau$ fiber.  Some of these additional topological features will appear in the solutions presented in this paper.

\subsection{As Euclidean-Einstein-Maxwell solutions}

One can show \cite{Bobev:2011kk, Bobev:2012af} that self-dual, harmonic 2-forms on LeBrun spaces can be written
\begin{equation}
\Theta \equiv \sum_{a=1}^3 \partial_a \bigg( \frac{H}{w} \bigg) \, \Op a = (\dd \tau + A) \wedge \dd \frac{H}{w} + w \hodge_3 \dd \frac{H}{w},
\end{equation}
where $H$ solves \eqref{lintoda} and $\hodge_3$ is taken with respect to the 3-metric
\begin{equation}
h = e^u (\dd x^2 + \dd y^2) + \dd z^2.
\end{equation}
By differentiating \eqref{toda} with respect to $z$, one can show that $u_z$ solves \eqref{lintoda}.  So define the Maxwell 2-form
\begin{equation}
\cF \equiv \Theta + \alpha J, \quad \text{with} \quad H = - \frac{u_z}{2 \alpha}.
\end{equation}
Then $(g, \cF)$ solve the Euclidean-Einstein-Maxwell equations \cite{LeBrun:2008kh}.

For simplicity in matching with the linear system found in \cite{Bena:2009fi}, we choose $\alpha = -1$, and hence
\begin{equation}
\Th 3 = \frac12 (\dd \tau + A) \wedge \dd \frac{u_z}{w} + \frac12 w \hodge_3 \dd \frac{u_z}{w}, \qquad \om 3 = J.
\end{equation}

\subsection{Floating branes on a LeBrun base}
\label{lebrun floating branes}

Next we solve the system \eqref{Z1 Th2 eqn}, \eqref{Z2 Th1 eqn}, \eqref{Z3 eqn}, \eqref{k eqn} on the LeBrun base.  We will find it convenient to define
\begin{equation}
K^3 \equiv \frac{u_z}{2}, \quad \text{such that} \quad \Th 3 = (\dd \tau + A) \wedge \dd \frac{K^3}{w} + w \hodge_3 \dd \frac{K^3}{w}.
\end{equation}
To solve the first layer, one makes the ans\"atze
\begin{align}
\Th{1} &= (\dd \tau + A) \wedge \dd \frac{K^1}{w} + w \hodge_3 \dd \frac{K^1}{w} + Z_2 \, (\Om{3} - \Op{3}), & Z_2 &= \frac{K^1 K^3}{w} + L_2, \label{Th1 Z2 ansatz} \\
\Th{2} &= (\dd \tau + A) \wedge \dd \frac{K^2}{w} + w \hodge_3 \dd \frac{K^2}{w} + Z_1 \, (\Om{3} - \Op{3}), & Z_1 &= \frac{K^2 K^3}{w} + L_1. \label{Th2 Z2 ansatz}
\end{align}
This leads to the linear equations
\begin{equation}
\partial_x^2 L_1 + \partial_y^2 L_1 + \partial_z^2 (e^u L_1) = 0, \qquad \partial_x^2 L_2 + \partial_y^2 L_2 + \partial_z^2 (e^u L_2) = 0, \label{L1 L2 eqns}
\end{equation}
and
\begin{align}
\partial_x^2 K^1 + \partial_y^2 K^1 + \partial_z (e^u \partial_z K^1) &= 2 \, \partial_z (e^u w \, L_2), \label{K1 eqn} \\
\partial_x^2 K^2 + \partial_y^2 K^2 + \partial_z (e^u \partial_z K^2) &= 2 \, \partial_z (e^u w \, L_1). \label{K2 eqn}
\end{align}
To solve the second layer, make the ans\"atze
\begin{equation}
k = \mu \, ( \dd \tau + A) + \omega, \qquad Z_3 = \frac{K^1 K^2}{w} + L_3, \qquad \mu = - \frac{K^1 K^2 K^3}{w^2} - \frac12 \sum_{I = 1}^3 \frac{K^I L_I}{w} + M. \label{k Z3 mu ansatz}
\end{equation}
Then the new functions $M$ and $L_3$ satisfy the equations
\begin{gather}
\partial_x^2 M + \partial_y^2 M + \partial_z (e^u \partial_z M) = \partial_z (e^u L_1 L_2), \label{M eqn} \\
\partial_x^2 L_3 + \partial_y^2 L_3 + e^u \, \partial_z^2 L_3 = 4 e^u w L_1 L_2 - 4 e^u w \, \partial_z M - 2 e^u (L_1 \, \partial_z K^1 + L_2 \, \partial_z K^2 ), \label{L3 eqn}
\end{gather}
and the 1-form $\omega$ satisfies
\begin{equation}
\begin{split}
\dd \omega &= w \hodge_3 \dd M - M \hodge_3 \dd w - u_z w M \hodge_3 \dd z - 2 w L_1 L_2 \hodge_3 \dd z \\
& \qquad + \frac12 \sum_I ( L_I \hodge_3 \dd K^I - K^I \hodge_3 L_I ) - \frac12 u_z (K^1 L_1 + K^2 L_2) \hodge_3 \dd z + \frac12 u_z K^3 L_3 \hodge_3 \dd z. \label{omega eqn}
\end{split}
\end{equation}
Therefore, to solve the ``floating brane" system on the LeBrun base, one first finds a function $u$ that solves the $SU(\infty)$ Toda equation, which also defines the function $K^3 \equiv \frac12 u_z$.  Then one solves \cref{lintoda,L1 L2 eqns,K1 eqn,K2 eqn,M eqn,L3 eqn}, in this order, for the seven remaining functions $w, K_1, K_2, L_1, L_2, L_3,$ and $M$.  Finally, one must solve \eqref{omega eqn} for the 1-form $\omega$.

\section{Axisymmetric K\"ahler base spaces}

Before we discuss solutions to the full system, we will explore the base space $g$ in detail.  Our task is to solve the $SU(\infty)$ Toda equation which, while known to be integrable, is also notoriously hard.  However, if we impose an additional $U(1)$ symmetry, there is a known method of attack \cite{Gaiotto:2009gz, Aharony:2012tz, ReidEdwards:2010qs, Santillan:1651, Ward:1990qt}.

First let us write the LeBrun metric in an explicitly $U(1) \times U(1)$-invariant form,
\begin{equation}
g = \frac{1}{w} (\dd \tau + A)^2 + w e^u \, (\dd r^2 + r^2 \, \dd \phi^2) + w \, \dd z^2,
\label{lebrun axi}
\end{equation}
where now all functions depend on $r, z$ only.  For completeness, the equations to be solved in these coordinates become
\begin{align}
\frac{1}{r} \partial_r (r u_r) + (e^u)_{zz} = 0, \label{toda axi} \\
\frac{1}{r} \partial_r (r w_r) + (e^u w)_{zz} = 0, \label{lintoda axi}
\end{align}
and
\begin{equation}
\dd A = r w_r \, \dd \phi \wedge \dd z + (e^u w)_z \, r \, \dd r \wedge \dd \phi.
\label{dA axi}
\end{equation}
At this point, we can solve \eqref{lintoda axi} and \eqref{dA axi} generically.  To accomplish this, note that the Laplacian on the 3-dimensional base $h$ is given by
\begin{equation}
e^u \laplace_h (\varphi) = \frac{1}{r} \partial_r (r \varphi_r) + (e^u \varphi_z)_z,
\end{equation}
and hence the Laplacian is related to the linearized Toda equation via $\partial_z$:
\begin{equation}
\partial_z \big( e^u \laplace_h (\varphi) \big) =  \frac{1}{r} \partial_r (r \partial_r \varphi_z) + (e^u \varphi_z)_{zz}.
\end{equation}
Therefore if we take some $\hat w$ which solves the Laplace equation on $h$
\begin{equation}
\frac{1}{r} \partial_r (r \hat w_r) + (e^u \hat w_z)_z = 0,
\label{w potential}
\end{equation}
then it is easy to show that \eqref{lintoda axi} and \eqref{dA axi} are solved by
\begin{equation}
w = \hat w_z, \qquad A = - r \hat w_r \, \dd \phi.
\end{equation}
One can think of $\hat w$ as a ``potential" that gives us the solutions for $w$ and $A$.

\subsection{Solving the axisymmetric Toda equation}

Now let us focus on the Toda equation with an axial symmetry \eqref{toda axi}.  The additional $U(1)$ symmetry allows one to make a B\"acklund transformation to new coordinates $\rho, \eta$ \cite{Gaiotto:2009gz, Aharony:2012tz, ReidEdwards:2010qs, Santillan:1651, Ward:1990qt}:
\begin{equation} \label{toda trans}
r^2 e^u = \rho^2, \qquad \log r = V_\eta, \qquad z = - \rho V_\rho.
\end{equation}
The Toda equation then reduces to the axisymmetric Laplace equation\footnote{Strictly speaking, this is a Poisson equation and we have ignored subtleties involving source terms (supported on a locus of measure zero) on the right-hand side of \eqref{toda axi}.  We avoid these subtleties by transforming the whole metric (taken as a local expression on an open chart) to the new coordinates $(\rho,\eta)$, while forgetting the old coordinates.  In Section \ref{boundary cond} we will discuss the source terms in the new coordinates which should appear in the right-hand-side of \eqref{V eqn}.  We remain agnostic about the exact form of the source terms as they would appear in the original coordinates \eqref{toda axi}, as this information is not necessary for constructing supergravity solutions.}  in $\RR^3$ in cylindrical coordinates:
\begin{equation} \label{V eqn}
\frac{1}{\rho} \partial_\rho (\rho V_\rho) + V_{\eta \eta} = 0.
\end{equation}
In principle, one must then invert the transformation \eqref{toda trans} to obtain $u$.  But in practice, for most functions $V$ this is intractable.  It is easier to change the metric to the new coordinates $\rho, \eta$, which results in
\begin{align} \label{lebrun metric cyl}
g &= \frac{1}{w} (\dd \tau + A)^2 + w \, h, \\
h &= \rho^2 (V_{\rho \eta}^2 + V_{\eta \eta}^2) (\dd \rho^2 + \dd \eta^2) + \rho^2 \, \dd \phi^2.
\end{align}
We should note that as a result of the transformations \eqref{toda trans}, the cylindrical coordinates $\rho, \eta, \phi$ inherit the orientation opposite to the usual:
\begin{equation}
\vol_h = \rho^2 (V_{\rho \eta}^2 + V_{\eta \eta}^2) \, \dd \rho \wedge \dd \eta \wedge \dd \phi.
\end{equation}

We must also change \eqref{lintoda axi} and \eqref{dA axi} into the new coordinates.  The Laplacian $\laplace_h$ becomes, up to an overall factor, the cylindrically-symmetric Laplacian on $\RR^3$,
\begin{equation}
\rho^2 (V_{\rho \eta}^2 + V_{\eta \eta}^2) \, \laplace_h(\varphi) = \frac{1}{\rho} \partial_\rho (\rho \varphi_\rho) + \varphi_{\eta \eta},
\end{equation}
and so the potential $\hat w$ solves
\begin{equation}
 \frac{1}{\rho} \partial_\rho (\rho \hat w_\rho) + \hat w_{\eta \eta} = 0,
\end{equation}
whose solutions we know well.  Then $w$ and $A$ are given by
\begin{equation} \label{w soln}
w = \hat w_z =  \frac{1}{\rho (V_{\rho \eta}^2 + V_{\eta \eta}^2)} \big( V_{\eta \eta} \, \hat w_\rho - V_{\rho \eta} \, \hat w_\eta \big).
\end{equation}
and
\begin{equation} \label{A soln}
A = - r \hat w_r \, \dd \phi = - \frac{1}{V_{\rho \eta}^2 + V_{\eta \eta}^2} \big( V_{\rho \eta} \, \hat w_\rho + V_{\eta \eta} \, \hat w_\eta \big) \, \dd \phi.
\end{equation}
Therefore, the geometric data of the base space are determined in terms of two functions $V, \hat w$ that solve the axisymmetric Laplace equation in $\RR^3$.

\subsection{Boundary conditions}
\label{boundary cond}

The task of writing an explicit base space is then reduced to solving cylindrically symmetric electrostatics problems in $\RR^3$ \cite{Gaiotto:2009gz}.  The question is what kinds of electrostatic problems give interesting solutions.

By analogy with BPS solutions on Gibbons-Hawking bases \cite{Bena:2007kg}, we expect to specify a collection of points along the $\eta$ axis where $w$ and $K^3 \equiv \frac12 u_z$ have poles.  The poles of $w$ control where the $\tau$ fiber pinches off, thus creating a series of homology 2-cycles (provided that the 3-dimensional base $h$ remain smooth at these points).  The poles of $u_z$ control sources of $\Th{3}$.  If $u_z$ has a pole where $w$ does not, we expect a singularity in the metric.  But if $u_z$ has poles coincident with poles of $w$, we expect that the base geometry be smooth (up to orbifold identifications), and such poles should control the fluxes of $\Th{3}$ on the adjacent 2-cycles.

In the simplest case, we consider where $w$ and $u_z$ each have a single, coincident pole.  Since both $w$ and $u_z$ solve the same elliptic linear PDE \cref{lintoda} (with the same boundary condition at infinity) and have only one ``source point'', it follows that $w$ and $u_z$ are proportional.  Hence $\Th{3} = 0$ and the metric is Ricci-flat (and therefore hyper-K\"ahler)---thus the metric \cref{lebrun axi} should be a Gibbons-Hawking metric, in alternative coordinates\footnote{In the general LeBrun ansatz, taking $w \sim u_z$ gives not a Gibbons-Hawking metric, but a more general hyper-K\"ahler manifold.  However, if we set $w \sim u_z$ in the $U(1) \times U(1)$-invariant ansatz of \cref{lebrun axi}, there is always some linear combination of the $U(1)$'s which is tri-holomorphic, hence the manifold must in fact be Gibbons-Hawking but written in unusual coordinates.}.  Looking at \eqref{lebrun axi}, we identify $z$ as the radial coordinate from the source point, and take $r, \phi$ to be stereographic coordinates on an $S^2$.  Hence we can write
\begin{equation} \label{uz near poles}
e^u = \frac{4 z^2}{(1 + r^2)^2}, \qquad u_z = \frac{2}{z}, \qquad w = \frac{q}{z},
\end{equation}
where $q$ is any integer.  Then as $z \to 0$, the metric \eqref{lebrun axi} is simply the flat metric on $\RR^4 / \ZZ_q$.  This gives the canonical example of coincident poles in $w, u_z$.  We expect that near any location where $w, u_z$ both blow up, the metric will locally have this form.

To get a function $u_z$ with many poles, we should choose a cylindrically-symmetric Laplace solution $V$ that gives rise to the behavior in \eqref{uz near poles}, and then use linearity to combine several solutions at centered at different points.  Using the B\"acklund transformation \cref{toda trans}, we have
\begin{equation} \label{uz condition}
u_z = \qquad \frac{2 V_{\eta \eta}}{\rho^2 (V_{\rho \eta}^2 + V_{\eta \eta}^2)} = - \frac{2}{\rho V_\rho} \qquad = \frac{2}{z},
\end{equation}
where the center equality is the boundary condition we need to satisfy near the source point in order for $u_z$ to have the appropriate singular behavior.  We see that while the cylindrically-symmetric Laplace equation for $V$ \eqref{V eqn} is linear, the boundary condition for $V$ is nonlinear.  To solve this boundary condition, one can guess a few known possibilities for $V$.  It turns out the appropriate choice is also the most obvious one to give a pole in the numerator:
\begin{equation}
V_{\eta \eta} = \frac{1}{\sqrt{\rho^2 + \eta^2}}.
\end{equation}
Integrating this twice with respect to $\eta$ and choosing appropriate integration constants, we find
\begin{equation}
V = - \sqrt{\rho^2 + \eta^2} + \eta \log \frac{\eta + \sqrt{\rho^2 + \eta^2}}{\rho}.
\end{equation}
Then we have
\begin{equation}
z = - \rho V_\rho = \sqrt{\rho^2 + \eta^2}, \qquad V_{\rho \eta} = - \frac{\eta}{\rho} \frac{1}{\sqrt{\rho^2 + \eta^2}},
\end{equation}
and hence
\begin{equation}
\rho^2 (V_{\rho \eta}^2 + V_{\eta \eta}^2) = 1, \quad \text{which implies} \quad u_z = \frac{2}{z},
\end{equation}
and the boundary condition is satisfied.  So we can write a solution with $N$ such poles as
\begin{align}
V &= k^3_0 \, \eta \log \rho + \sum_{i=1}^N k^3_i \, H_i(\rho, \eta), \label{V def} \\
H_i(\rho, \eta) &= - \sqrt{\rho^2 + (\eta - \eta_i)^2} + (\eta - \eta_i) \log \frac{\eta - \eta_i + \sqrt{\rho^2 + (\eta - \eta_i)^2}}{\rho},
\end{align}
where $\eta_i$ are the locations of the poles on the $\eta$ axis.

\begin{figure}
\centering
\includegraphics[height=4.2cm]{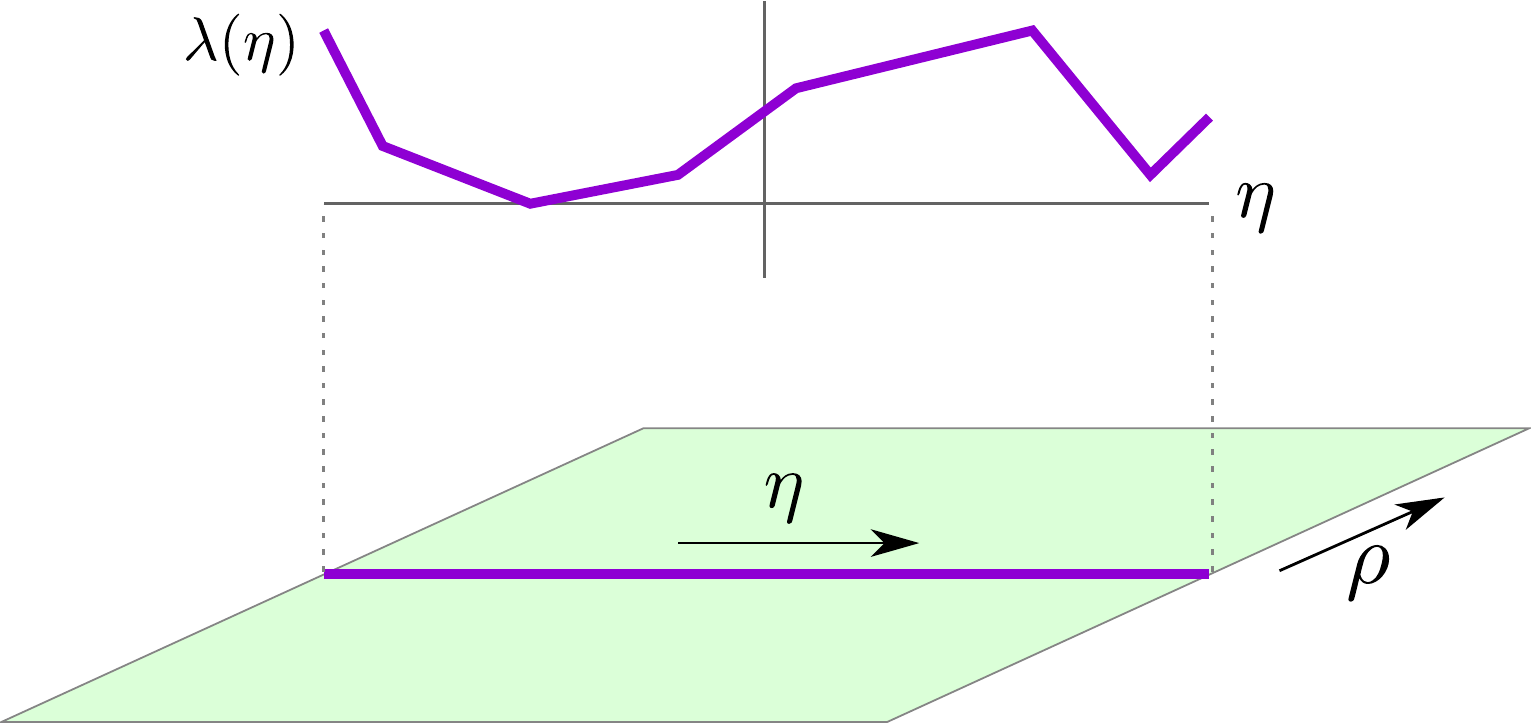}
    \caption{\it \small The electrostatics problem corresponding to $V$.  $\lambda(\eta)$ is a line charge density profile along the $\eta$ axis, which is piecewise linear with ``kinks" at each of the $\eta_i$.}
\label{kinks}
\end{figure}

Interpreted as an electrostatics problem, this corresponds to the potential of a line charge along the $\eta$ axis of varying charge density $\lambda(\eta)$.  The charge density profile $\lambda(\eta)$ is piecewise linear, with a ``kink" at each $\eta_i$ as in Fig. \ref{kinks}, such that
\begin{equation}
\lambda''(\eta) = \sum_{i = 1}^N k^3_i \, \delta(\eta - \eta_i).
\end{equation}
where the parameters $k^3_i$ represent the amount by which the slope jumps as one moves across the kink at $\eta_i$:
\begin{equation}
k^3_i \equiv \frac{d \lambda}{d \eta} \Big|_{\eta_i + \epsilon} - \frac{d \lambda}{d \eta} \Big|_{\eta_i - \epsilon}.
\end{equation}
In $V$ \cref{V def}, we have also put an additional parameter $k^3_0$, which represents the freedom to choose the value of $\lambda'(\eta)$ at infinity\footnote{Specifically, $2 \, k^3_0$ is the sum $\lambda'(\infty) + \lambda'(-\infty)$, while the difference $\lambda'(\infty) - \lambda'(-\infty)$ is given by the sum of all the jumps $k^3_i$.}.

\begin{figure}
\centering
\includegraphics[height=4.2cm]{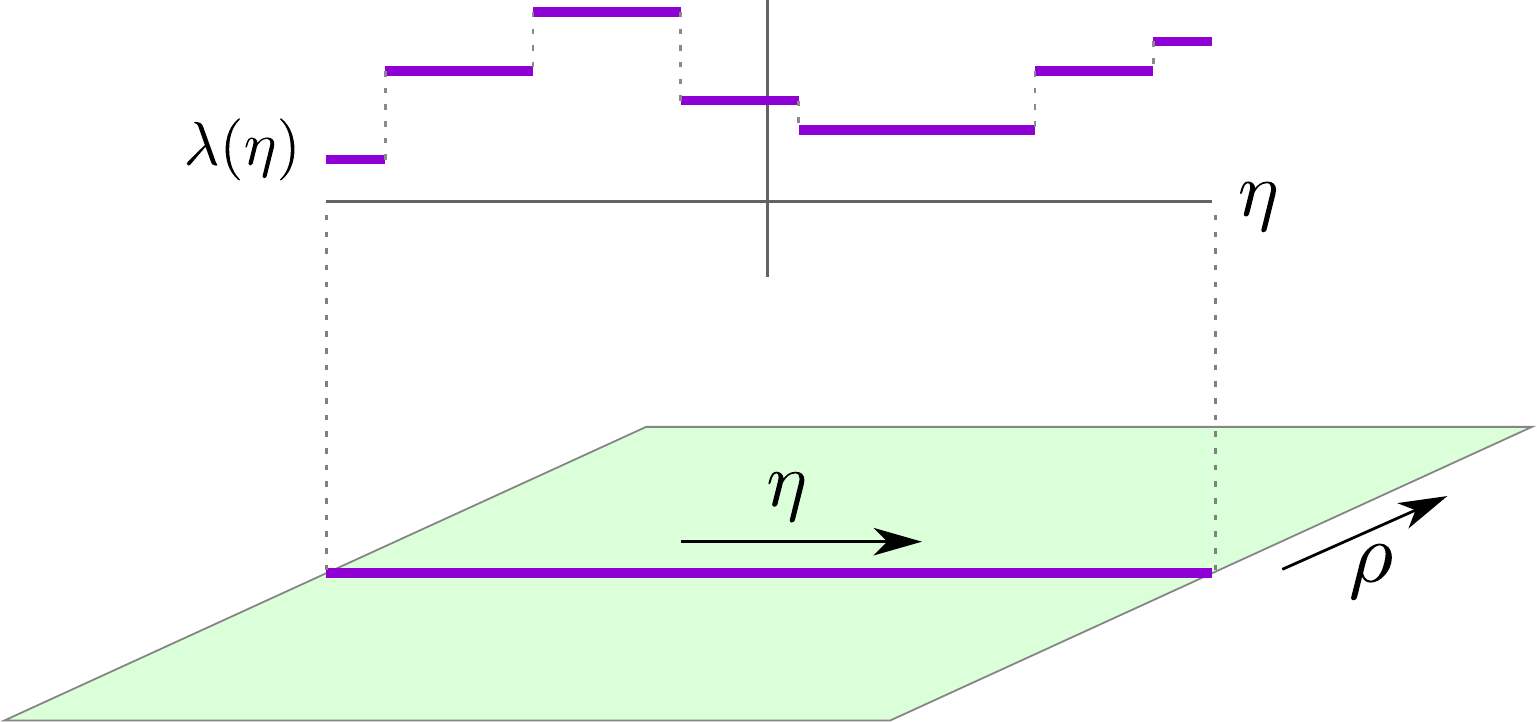}
    \caption{\it \small The electrostatics problem corresponding to $\hat w$.  The line charge profile $\lambda(\eta)$ is piecewise constant, with ``jumps" at each $\eta_i$.}
\label{jumps}
\end{figure}

We also choose $\hat w$ such that $w = \hat w_z$ has $1/z$ type behavior at the Gibbons-Hawking points.  It is easy to show that correct choice is
\begin{align}
\hat w &= q_0 \log \rho + \sum_{i=1}^N q_i \, G_i(\rho, \eta), \label{w hat def} \\
G_i(\rho, \eta) &= \log \frac{\eta - \eta_i + \sqrt{\rho^2 + (\eta - \eta_i)^2}}{\rho}.
\end{align}
As an electrostatics problem, this corresponds to a line charge profile $\lambda(\eta)$ which is piecewise constant, with ``jumps" at each $\eta_i$ as in Fig. \ref{jumps}.  The parameters $q_i$ give the amount of each jump:
\begin{equation}
q_i \equiv \lambda(\eta) \Big|_{\eta_i + \epsilon} - \lambda(\eta) \Big|_{\eta_i - \epsilon},
\end{equation}
(where this $\lambda(\eta)$ is the one in Fig. \ref{jumps}).

For completeness, it is helpful to write out the $\eta$- and $\rho$-derivatives of these, which  appear in all other formulas:
\begin{align}
V_{\eta \eta} &= \sum_{i=1}^N \frac{k^3_i}{\Sigma_i}, & V_{\rho \eta} &= \frac{k^3_0}{\rho} - \frac{1}{\rho} \sum_{i=1}^N \frac{k^3_i \, (\eta - \eta_i)}{\Sigma_i}, \label{V der def} \\
\hat w_\eta &= \sum_{i=1}^N \frac{q_i}{\Sigma_i}, & \hat w_\rho &=  \frac{q_0}{\rho} - \frac{1}{\rho} \sum_{i=1}^N \frac{q_i \, (\eta - \eta_i)}{\Sigma_i}, \label{wh der def}
\end{align}
where $\Sigma_i \equiv \sqrt{\rho^2 + (\eta - \eta_i)^2}$.  We note that this is essentially the same construction as in \cite{Abreu2009} for scalar-flat toric K\"ahler 4-manifolds (which can always be written in LeBrun form).  Thus the base space is defined via the functions \cref{V der def,wh der def} and the $2N+2$ parameters $k^3_0, k^3_i, q_0, q_i$.

\subsection{Near the singularities}
\label{near points}

The base space is constructed out of $N$ points where the functions $V$ and $w$ are singular.  In this section we look in the neighborhood of these points and show that the manifold is perfectly smooth, up to orbifold identifications, in a similar manner to Gibbons-Hawking metrics \cite{Gibbons:1979zt}.  Specifically we will find that the metric \cref{lebrun metric cyl} at these points locally approaches the orbifold $\RR^4 / G$, where $G \simeq \ZZ_m \times \ZZ_n$ is a finite subgroup of the maximal torus\footnotemark{} $U(1) \times U(1) \subset SO(4)$.

\footnotetext{We note that the factors $\ZZ_m, \ZZ_n \subset U(1) \times U(1)$ are not necessarily rotations \emph{in a plane} (i.e. fixing every point in the orthogonal plane).  One can have, for example, $\ZZ_m$ acting in the first $U(1)$ and $\ZZ_n$ acting in the diagonal $U(1)$.  Rotations in the diagonal $U(1)$ fix only the origin.}

Taking the limit as $(\rho, \eta) \to (0, \eta_\ell)$ for some $\eta_\ell$, we can define new coordinates
\begin{equation}
\rho = R \sin \theta, \qquad \eta - \eta_\ell = R \cos \theta.
\end{equation}
We will find it convenient to define the quantities
\begin{equation}
\bar K^3_\ell \equiv \sum_{\substack{i \\ i \neq \ell}} k^3_i \sign (\eta_\ell - \eta_i), \qquad \bar Q_\ell \equiv \sum_{\substack{i \\ i \neq \ell}} q_i \sign (\eta_\ell - \eta_i),
\end{equation}
and also the functions
\begin{align}
\widetilde{K}(\theta) &\equiv (k^3_\ell)^2 + (\bar K^3_\ell - k^3_0)^2 + 2 \, k^3_\ell (\bar K^3_\ell - k^3_0) \cos \theta, \\
\widetilde{Q}(\theta) &\equiv q_\ell^2 + (\bar Q_\ell - q_0)^2 + 2 \, q_\ell (\bar Q_\ell - q_0) \cos \theta, \\
\widetilde{KQ}(\theta) &\equiv k^3_\ell q_\ell + (\bar K^3_\ell - k^3_0) (\bar Q_\ell - q_0) + \big( k^3_\ell (\bar Q_\ell - q_0) + q_\ell (\bar K^3_\ell - k^3_0) \big) \cos \theta.
\end{align}
Then for small $R$, we have
\begin{equation} \label{near points}
\rho^2 (V_{\eta \eta}^2 + V_{\rho \eta}^2) \to \widetilde{K}(\theta), \qquad w \to \frac{1}{\widetilde{K}(\theta)} \frac{\widetilde{q}_\ell}{R}, \qquad A \to - \frac{\widetilde{KQ}(\theta)}{\widetilde{K}(\theta)} \, \dd \phi,
\end{equation}
where we define the determinant:
\begin{equation}
\widetilde{q}_\ell \equiv q_\ell (\bar K^3_\ell - k^3_0) - k^3_\ell (\bar Q_\ell - q_0).
\label{qt def}
\end{equation}

The metric becomes
\begin{equation} \label{metric near points}
\dd s^2 = \frac{\widetilde{K}(\theta) R}{\widetilde{q}_\ell} \bigg( \dd \tau - \frac{\widetilde{KQ}(\theta)}{\widetilde{K}(\theta)} \, \dd \phi \bigg)^2 + \frac{\widetilde q_\ell}{R} \big( \dd R^2 + R^2 \, \dd \theta^2 \big) + \frac{\widetilde{q}_\ell R}{\widetilde{K}(\theta)} \sin^2 \theta \, \dd \phi^2,
\end{equation}
which, surprisingly enough, is flat.  Setting $R = \varrho^2 / (4 \, \widetilde q_\ell)$, this can be rearranged into the more convenient form
\begin{equation}
\dd s^2 = \dd \varrho^2 + \frac{\varrho^2}{4} \bigg[ \dd \theta^2 + \frac{1}{\widetilde q_\ell{}^2} \bigg( \widetilde{K}(\theta) \, \dd \tau^2 - 2 \widetilde{KQ}(\theta) \, \dd \tau \, \dd \phi + \widetilde{Q}(\theta) \, \dd \phi^2 \bigg) \bigg].
\label{source point metric}
\end{equation}
We compare this to a flat metric\footnote{This metric is related to the standard spherical coordinates on $\RR^4$ by $\theta = 2 \vartheta$.} on $\RR^4$:
\begin{equation}
\dd s^2 = \dd \varrho^2 + \frac{\varrho^2}{4} \bigg[ \dd \theta^2 + 2 \big( 1 + \cos \theta \big) \, \dd \alpha^2 + 2 \big( 1 - \cos \theta \big) \, \dd \beta^2 \bigg],
\label{R4 metric}
\end{equation}
where both $\alpha, \beta$ are (ordinarily) identified modulo $2 \pi$ and $\theta \in [0, \pi]$.  The metrics \cref{source point metric} and \cref{R4 metric} are then related by a coordinate transformation
\begin{align}
\tau &= (q_\ell - \bar Q_\ell + q_0) \, \alpha - (q_\ell + \bar Q_\ell - q_0) \, \beta, \label{tau trans} \\
\phi &= (k^3_\ell - \bar K^3_\ell + k^3_0) \, \alpha - (k^3_\ell + \bar K^3_\ell - k^3_0) \, \beta. \label{phi trans}
\end{align}

To discover the precise geometry in the neighborhood of the origin, we must carefully follow the identifications of the angular coordinates.  It is natural to identify the coordinates $(\tau, \phi)$ on the ``diamond'' lattice $\Gamma_{LB}$, given by the identifications
\begin{equation} \label{tau lattice}
(\tau, \phi) : \quad (0,0) \sim (4\pi,0) \sim (2\pi,2\pi) \sim (2\pi,-2\pi),
\end{equation}
whose basis can be written as a matrix $\Lambda_{LB}$ of column vectors which represent the coordinates where $(\tau, \phi)$ are identified:
\begin{equation}
\Lambda_{LB} = 2 \pi \begin{pmatrix} 1 & 1 \\ 1 & -1 \end{pmatrix}, \quad \text{or} \quad \Lambda_{LB} = 2 \pi \begin{pmatrix} 2 & 1 \\ 0 & 1 \end{pmatrix}.
\end{equation}
We are free to choose any pair of column vectors that generate the same lattice of identifications; alternatively, $\Lambda_{LB}$ is defined only up to right action by $GL(2,\ZZ)$\footnote{We define $GL(2,\ZZ)$ as the group of $2 \times 2$ matrices with integer entries and determinant $\pm 1$, hence invertible over $\ZZ$.  This group is sometimes also called $S^*\!L(2,\ZZ)$ or $SL^\pm(2,\ZZ)$.}.  Then applying the coordinate transformation \cref{tau trans,phi trans}, we find that the $(\alpha, \beta)$ coordinates should be identified on the lattice $\widetilde \Gamma$, generated by the basis
\begin{equation} \label{alpha lattice}
\widetilde \Lambda = 2\pi \cdot \frac{1}{2 \widetilde q_\ell}
\begin{pmatrix}
k^3_\ell + \hat K^3_\ell + q_\ell + \hat Q_\ell \; & k^3_\ell + \hat K^3_\ell - q_\ell - \hat Q_\ell \\
k^3_\ell - \hat K^3_\ell + q_\ell - \hat Q_\ell \; & k^3_\ell - \hat K^3_\ell - q_\ell + \hat Q_\ell
\end{pmatrix},
\end{equation}
where for ease of legibility we have defined
\begin{equation}
\hat K^3_\ell \equiv \bar K^3_\ell - k^3_0, \qquad \hat Q_\ell \equiv \bar Q_\ell - q_0.
\end{equation}
We should then compare this lattice to a ``reference'' lattice $\Gamma$, generated by the basis
\begin{equation} \label{std lattice}
\Lambda = 2\pi \begin{pmatrix} 1 & 0 \\ 0 & 1 \end{pmatrix},
\end{equation}
which represents the ordinary $2\pi$ identifications that $(\alpha, \beta)$ would take if there were no conical singularity.  In order for the LeBrun metric to approach a proper \emph{orbifold} $\RR^4 / G$ at the source point, one requires that the lattices $\Gamma, \widetilde \Gamma$ be compatible---that is, one must have that $\Gamma$ is a sublattice of $\widetilde \Gamma$.  Otherwise, one has a conical point that is not an orbifold\footnote{An analogous situation with orbifolds of $\RR^2$ is that the angular coordinate must be identified modulo $2\pi/n$, but not modulo $2\pi m / n$ for some $m > 1$ ($m,n$ relatively prime), as this would fail to be a quotient.}.

The condition that $\Gamma \subseteq \widetilde \Gamma$ as lattices is equivalent to requiring that $\widetilde \Lambda^{-1} \Lambda$ be an \emph{integer} matrix.  We want this to be true at every $\eta_\ell$, in principle giving $N$ conditions; however, all of these conditions are equivalent to a single parity condition:
\begin{equation} \label{even cond}
\bigg( k^3_0 + \sum_{i=1}^N k^3_i + q_0 + \sum_{i=1}^N q_i \bigg) \in 2\ZZ,
\end{equation}
that is, the sum of all the parameters $k^3_0, k^3_i, q_0, q_i$ must be even.  If we impose this condition, then at every $\eta_\ell$ the metric will approach an orbifold singularity.

At a given $\eta_\ell$, we can then compute the group $G$ that describes this orbifold singularity.  The details are given in Appendix \ref{orbifold groups}.  The general procedure is as follows:  Given the lattices $\widetilde \Gamma, \Gamma$ generated by \cref{alpha lattice,std lattice}, one can find the group $G$ by reducing $\widetilde \Lambda^{-1} \Lambda$ to \emph{Smith normal form}, where one diagonalizes $\widetilde \Lambda^{-1} \Lambda$ by left and right $GL(2,\ZZ)$ actions:
\begin{equation} 
R = \widetilde P^{-1} \widetilde \Lambda^{-1} \Lambda P, \qquad R = \begin{pmatrix} r_1 & 0 \\ 0 & r_2 \end{pmatrix}, \quad \text{where} \quad P, \widetilde P \in GL(2,\ZZ).
\end{equation}
Given the parity condition \cref{even cond}, it is always true that $\widetilde \Lambda^{-1} \Lambda = 2 \pi \widetilde \Lambda^{-1}$ has integer entries.  Then the numbers $r_1, r_2$ are integers, and determine $G$ via
\begin{equation} \label{G mn}
G = \ZZ_m \times \ZZ_n, \quad \text{where} \quad m = r_1, \quad n = r_2.
\end{equation}

\subsubsection{Specific details of the groups $G$}

We then find a number of interesting facts (whose detailed derivation can be found in \cref{orbi group results}).

First, at every orbifold point one has, as mentioned, that $\Gamma \subseteq \widetilde \Gamma$ as a sublattice, and the group $G$ is formally given by the quotient $G \simeq \widetilde \Gamma / \Gamma$.  The order of the group $G$ is
\begin{equation} \label{G order}
\# G ~=~ \abs{\det(\widetilde \Lambda^{-1} \Lambda)} ~=~ \abs{\det(2\pi \widetilde \Lambda^{-1})} ~=~ \abs{\widetilde q_\ell},
\end{equation}
and thus the group $G$ is trivial exactly when $\widetilde q_\ell = \pm 1$.  At such points, the metric approaches flat $\RR^4$ with no conical singularity.

Second, we can ask when the orbifold point at $\eta_\ell$ is similar to the orbifold point of a charge $m > 1$ Gibbons-Hawking metric.  These are points where $G \simeq \ZZ_m$ and the action of $\ZZ_m$ is in the diagonal $U(1)$ of the maximal torus $U(1) \times U(1) \in SO(4)$.  We find that such orbifold points occur whenever:
\begin{equation} \label{diag cond}
\widetilde q_\ell = \pm m, \qquad \frac{2 (\bar K^3_\ell - k^3_0)}{\widetilde q_\ell} \in \ZZ, \qquad \text{and} \qquad \frac{2(\bar Q_\ell - q_0)}{\widetilde q_\ell} \in \ZZ.
\end{equation}
One can also consider $G \simeq \ZZ_m$ acting in the \emph{anti}-diagonal $U(1)$, which results in similar conditions:
\begin{equation} \label{anti diag cond}
\widetilde q_\ell = \pm m, \qquad \frac{2 \, k^3_\ell}{\widetilde q_\ell} \in \ZZ, \qquad \text{and} \qquad \frac{2 \, q_\ell}{\widetilde q_\ell} \in \ZZ.
\end{equation}

More generally, $G \simeq \ZZ_m \times \ZZ_n$ where each $\ZZ_k$ acts in some linear combination of the two $U(1)$'s.  In the simplest case, the $\ZZ_k$ act by rotation within a plane; i.e. by rotating $(x^1, x^2)$ and leaving $(x^3, x^4)$ fixed.  However, the ``diagonal'' rotations discussed above act in both planes and do not fix any point aside from the origin.  One can also obtain more general rotations that rotate both $(x^1, x^2)$ and $(x^3, x^4)$ planes by unequal amounts.

In any case, an orbifold singularity with a finite group action such as $\RR^4 / G$ is benign in string theory \cite{Aspinwall:1994ev}, so in the context of microstate geometries, we will count such points as regular.

\subsection{At infinity}
\label{base asym}

In the asymptotic region of the base metric, let us define
\begin{equation}
\rho = R \sin \theta, \qquad \eta = R \cos \theta.
\end{equation}
Then as $R \to \infty$, we have
\begin{align}
\rho^2 (V_{\rho \eta}^2 + V_{\eta \eta}^2) &\to (k^3_0)^2 + (K^3_\star)^2 - 2 \, k^3_0 K^3_\star \cos \theta, \label{r2U asym} \\
w &\to \bigg( \frac{q_0 K^3_\star - k^3_0 Q_\star}{(k^3_0)^2 + (K^3_\star)^2 - 2 \, k^3_0 K^3_\star \cos \theta} \bigg) \frac{1}{R}, \label{w asym} \\
A &\to \bigg( \frac{k^3_0 q_0 + K^3_\star Q_\star - (q_0 K^3_\star + k^3_0 Q_\star) \cos \theta}{(k^3_0)^2 + (K^3_\star)^2 - 2 \, k^3_0 K^3_\star \cos \theta} \bigg) \dd \phi, \label{A asym}
\end{align}
where the quantities $K^3_\star, Q_\star$ are defined as
\begin{equation}
K^3_\star \equiv \sum_{i=1}^N k^3_i, \qquad Q_\star \equiv \sum_{i=1}^N q_i.
\end{equation}
We see that \cref{r2U asym,w asym,A asym} have the same structure as \cref{near points}.  So at infinity, the base metric approaches a metric with the same structure as \cref{metric near points}.  We can define the determinant
\begin{equation} \label{q  infty def}
\widetilde q_\infty \equiv q_0 K^3_\star - k^3_0 Q_\star,
\end{equation}
and then the conditions \cref{G order} and \cref{diag cond}, \cref{anti diag cond} apply in the same way.  In particular, one has smooth $\RR^4$ at infinity whenever
\begin{equation} \label{trivial cond infty}
\widetilde q_\infty = \pm 1.
\end{equation}
One can obtain $\RR^4 / \ZZ_m$, where $\ZZ_m$ acts on the diagonal $U(1)$ via
\begin{equation} \label{diag cond infty}
\widetilde q_\infty = \pm m, \qquad \frac{2 \, K^3_\star}{\widetilde q_\infty} \in \ZZ, \qquad \text{and} \qquad \frac{2 \, Q_\star}{\widetilde q_\infty} \in \ZZ,
\end{equation}
or where $\ZZ_m$ acts on the anti-diagonal $U(1)$ via
\begin{equation} \label{anti diag cond infty}
\widetilde q_\infty = \pm m, \qquad \frac{2 \, k^3_0}{\widetilde q_\infty} \in \ZZ, \qquad \text{and} \qquad \frac{2 \, q_0}{\widetilde q_\infty} \in \ZZ.
\end{equation}
In general, the geometry approaches $\RR^4 / G_\infty$, where again $G_\infty \simeq \ZZ_m \times \ZZ_n$.

\subsection{Ambipolar bases}
\label{ambipolar}

If the base space is considered in isolation, then we must restrict the ``charges'' $\widetilde q_\ell$ at each point to be positive.  Otherwise, the function $w$ will change sign\footnote{Caveat:  This is not quite true, as we will show in \cref{eng soln}.}, and the signature of the metric \cref{lebrun axi} will flip from $\sig40$ to $\sig04$.

However, in the context of supergravity solutions, the metric \cref{lebrun axi} appears multiplied by the warp factor $Z = (Z_1 Z_2 Z_3)^{1/3}$ in the full 5-dimensional metric,
\begin{equation}
\dd s_5^2 = - Z^{-2} \, (\dd t + k)^2 + Z \, \dd s_4^2.
\end{equation}
Therefore, we can allow $w$ to change sign, so long as each of the $Z_1, Z_2, Z_3$ changes sign along the same locus, such that the 5-dimensional metric retains the signature $\sig41$.  We call such a base space ``ambipolar'', where the signature is allowed to flip from $\sig40$ to $\sig04$, as has been discussed at length in \cite{Bena:2007kg, Gibbons:2013tqa}.  This justifies the use of $\widetilde q_\ell, \widetilde q_\infty = \pm 1, \pm m$ in \cref{diag cond,anti diag cond,trivial cond infty,diag cond infty,anti diag cond infty}.

With this allowed flexibility in the charges $\widetilde q_\ell$, we can construct a wide variety of base spaces.  In particular, it should be possible to have both $\widetilde q_\ell = \pm 1$ at every point \emph{and} $\widetilde q_\infty = \pm 1$ at infinity, thus allowing us to write down supergravity solutions with an arbitrary number of bubbles and no orbifold points anywhere.

\subsection{Engineering solutions}
\label{eng soln}

Here we will describe a simple algorithm for generating solutions with an arbitrary number of points $\eta_\ell$, each of which has trivial orbifold group (and thus is smooth).  We will assume that each $\widetilde q_\ell = +1$ in order to show an interesting result.  It is simple to generalize this algorithm to the more flexible ambipolar case where $\widetilde q_\ell = \pm 1$.

To derive this algorithm, we first observe that
\begin{equation}
\bar Q_{i+1} - \bar Q_i = q_i + q_{i+1},
\end{equation}
and hence one has
\begin{equation} \label{recur}
(\bar Q_{i+1} + q_{i+1}) = (\bar Q_i + q_i) + 2 q_{i+1},
\end{equation}
and similarly for $\bar K^3_i$.  The parity condition \cref{even cond} can also be written
\begin{equation} \label{parity}
k^3_0 + q_0 + (\bar Q_i + q_i) + (\bar K^3_i + k^3_i) \in 2 \ZZ,
\end{equation}
where $i \in \{1\ldots N\}$ is any of the $N$ points.  Since the $q_i$ are integers, \cref{recur} guarantees that if \cref{parity} is true for any given $i$, it is true for all $i$.  Therefore without explicitly writing down the sum of all the parameters, we can describe a recursive algorithm for constructing solutions starting at $i=1$ and adding as many points as we like.

A second observation we will need is that
\begin{align}
\widetilde q_{i+1} &\equiv q_{i+1} (\bar K^3_{i+1} - k^3_0) - k^3_{i+1} (\bar Q_{i+1} - q_0) \\
&= q_{i+1} (\bar K^3_{i+1} + k^3_{i+1} - k^3_0) - k^3_{i+1} (\bar Q_{i+1} + q_{i+1} - q_0) \\
\begin{split} \label{recur 2}
&= q_{i+1} (\bar K^3_i + k^3_i + 2k^3_{i+1} - k^3_0) \\
& \qquad \qquad \qquad \qquad - k^3_{i+1} (\bar Q_{i} + q_{i} + 2q_{i+1} - q_0)
\end{split} \\
\widetilde q_{i+1} &= q_{i+1} (\bar K^3_i + k^3_i - k^3_0) - k^3_{i+1} (\bar Q_{i} + q_{i} - q_0), \label{recur q}
\end{align}
where the third line \cref{recur 2} follows from \cref{recur}.  Since we wish to set each $\widetilde q_i = 1$, the last line \cref{recur q} gives us a recurrence relation for the parameters $q_i, k^3_i$.  Then the algorithm proceeds as follows:
\begin{enumerate}
\item Define
\begin{equation} \label{ab def}
a_i \equiv \bar K^3_i + k^3_i - k^3_0, \qquad b_i \equiv \bar Q_{i} + q_{i} - q_0,
\end{equation}
and choose any $a_1, b_1, k^3_1, q_1$ such that
\begin{equation}
\widetilde q_1 \equiv q_1 \, a_1 - k^3_1\,  b_1 = 1, \qquad a_1 + b_1 + k^3_1 + q_1 \in 2\ZZ.
\end{equation}

\item Next, find some $k^3_2, q_2$ such that (using \cref{recur q})
\begin{equation}
\widetilde q_2 = q_2 \, a_1 - k^3_2 \, b_1 = 1,
\end{equation}
\emph{and} such that
\begin{equation}
a_2 = a_1 + 2 \, k^3_2, \qquad b_2 = b_1 + 2 \, q_2
\end{equation}
are relatively prime\footnote{This is required in order for the next constraint $\widetilde q_{i+1} = 1$ to have a solution.}.

\item Repeat this as many times as desired, finding some $k^3_{i+1}, q_{i+1}$ such that
\begin{equation}
\widetilde q_{i+1} = q_{i+1} \, a_i - k^3_{i+1} \, b_i = 1,
\end{equation}
and
\begin{equation}
a_{i+1} = a_i + 2 \, k^3_{i+1}, \qquad b_{i+1} = b_i + 2 \, q_{i+1}
\end{equation}
are relatively prime.

\item After choosing $N$ such $k^3_i, q_i$, plug them all back into the definitions \cref{ab def} along with $a_1, b_1$ from the initial step, and solve for the remaining parameters $k^3_0, q_0$.
\end{enumerate}
It is simple to generalize this algorithm to produce a sequence of points with any desired $\widetilde q_i$.  In this case, the requirement that each $a_i, b_i$ be relatively prime can be weakened, noting that in general, $\gcd(a_i, b_i)$ must divide both $\widetilde q_i$ and $\widetilde q_{i+1}$.

We also note that in the final step of the algorithm, there is no longer any freedom to choose parameters, and $k^3_0, q_0$ must be solved for, from \cref{ab def}.  Therefore once we have laid down a sequence of $N$ points with given $\widetilde q_i$, the orbifold structure at infinity is fixed\footnote{However, the orbifold structure at infinity depends on the specific $k^3_i, q_i$ of the solution, and the same sequence of $\widetilde q_i$ can result in different asymptotics!}.

If a specific behavior at infinity is required, one can re-write the algorithm to work backwards.  The ``reverse'' algorithm is \emph{not} identical to the one written here, but it is simple to work out from the reasoning in \cref{recur,parity} along similar lines.


Using this algorithm it is easy to obtain some interesting solutions.  We will give only the solutions and not the details of the algorithm used to obtain them.  These two examples show some surprising features which emphasize the difference between LeBrun metrics and Gibbons-Hawking metrics regarding the types of allowed orbifold points:

\subsubsection{Example 1: Every interior $\widetilde q_i = 1$, but at infinity $\widetilde q_\infty = -1$}

The first example has three points, and is given by the parameters:
\begin{align}
q_1 &= 4, & q_2 &= -3, & q_3 &= 2; & q_0 &= -2, \\
k^3_1 &= 5, & k^3_2 &= -4, & k^3_3 &= 1; & k^3_0 &= -1.
\end{align}
For this example, one has
\begin{equation}
\widetilde q_1 = 1, \qquad \widetilde q_2 = 1, \qquad \widetilde q_3 = 1, \qquad \widetilde q_\infty = -1.
\end{equation}
Hence at all the source points $\eta_i$ one has smooth $\RR^4$ with trivial orbifold group.  However, the minus sign in $\widetilde q_\infty$ reveals that it is possible for a LeBrun metric to flip signature $\sig40$ to $\sig04$ at infinity even if all the interior points have positive ``charges''!

This also implies that the na\"ive positivity condition mentioned at the beginning of \cref{ambipolar} is not quite correct, and requires that one also take into account the numerator of \cref{w asym} to have a metric with positive signature everywhere.  Since in the context of higher-dimensional supergravity solutions we do not require the signature of the base to remain $\sig40$ everywhere, we will not worry about this.

\subsubsection{Example 2: Every interior $\widetilde q_i \geq 1$, but at infinity $\widetilde q_\infty = +1$}

A second important example is also given by three points:
\begin{align}
q_1 &= -1, & q_2 &= 2, & q_3 &= 2; & q_0 &= 2, \label{asym flat q} \\
k^3_1 &= 0, & k^3_2 &= 1, & k^3_3 &= 1; & k^3_0 &= 1. \label{asym flat k}
\end{align}
and this example has
\begin{equation}
\widetilde q_1 = 3, \qquad \widetilde q_2 = 1, \qquad \widetilde q_3 = 1, \qquad \widetilde q_\infty = 1.
\end{equation}
In this case the metric does not unexpectedly flip signature.  However, we do see that it is possible for a LeBrun metric to be asymptotically \emph{flat} (and not just locally flat) even if the interior ``charges'' are all positive and some of them are greater than 1.  This is in contrast to Gibbons-Hawking metrics, where it is a mathematical theorem that the only asymptotically (globally) flat hyper-K\"ahler metric in 4 dimensions is $\RR^4$ \cite{Gibbons:1979xn}.  Because LeBrun metrics are merely K\"ahler and not hyper-K\"ahler, they are not subject to this restriction, and the set of parameters \cref{asym flat q,asym flat k} give an explicit example to this effect.


It does not, however, appear to be possible to choose parameters such that all the $\widetilde q_i = +1$ \emph{and} $\widetilde q_\infty = +1$, although we have not found a way to prove this impossibility in general.

\subsection{A topological m\'enagerie}

We have shown that the base metric approaches $\RR^4 / G$, for $G \simeq \ZZ_m \times \ZZ_n$, near each of the geometric charges where the $\tau$ fiber pinches off.  As explained in \cref{lebrun topo}, these points control the appearance of homology 2-spheres as the $\tau$ fiber sweeps along a path between any two such points.

There are also additional phenomena which appear when we look more carefully at the axis in the 3-dimensional base $h$:
\begin{equation}
\rho^2 (V_{\rho \eta}^2 + V_{\eta \eta}^2) (\dd \rho^2 + \dd \eta^2) + \rho^2 \, \dd \phi^2.
\end{equation}
Along the axis, but away from the Gibbons-Hawking points, one has
\begin{equation} \label{a def}
\rho^2 (V_{\rho \eta}^2 + V_{\eta \eta}^2) \to \bigg( k^3_0 - \sum_{i=1}^N k^3_i \sign (\eta - \eta_i) \bigg)^2 \equiv a^2,
\end{equation}
which is a piecewise-constant function with jumps at each $\eta_i$.  Whenever $a^2 = 1$, then as $\rho \to 0$, the $\phi$ circle pinches off smoothly.  If instead $a^2 \neq 1$ and $a^2 > 0$, then the $\phi$ circle pinches off in a conical singularity $\RR^2 / \ZZ_a$.

\begin{figure}
\centering
\includegraphics[height=3cm]{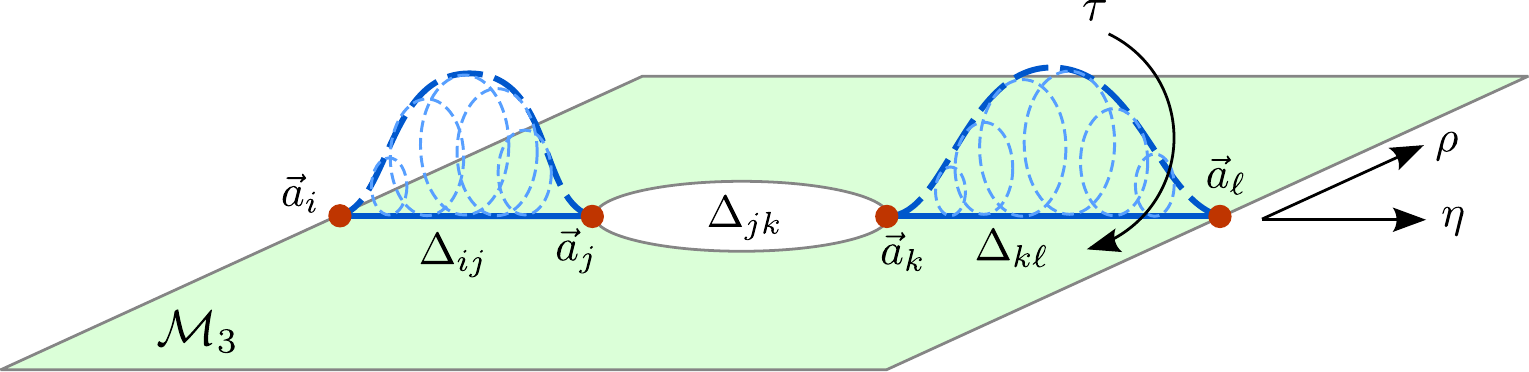}
    \caption{\it \small Homology 2-cycles in the axisymmetric base space.  $\Delta_{ij}$ and $\Delta_{k\ell}$ are cycles formed by sweeping the $\tau$ fiber between source points.  $\Delta_{jk}$ is a cycle formed by the $\phi$ circle.  In the $\rho, \eta$ coordinates, the $\phi$-cycle appears as a line segment between $\vec a_j$ and $\vec a_k$.  However, $\phi$ does not pinch off there, but approaches a finite size as $\rho \to 0$.} 
\label{phi cycles}
\end{figure}

But it is also possible that $a = 0$.  Expanding to the next order in $\rho^2$, and imposing
\begin{equation}
k^3_0 = \sum_{i=1}^N k^3_i \sign (\eta - \eta_i),
\end{equation}
one has, as $\rho \to 0$,
\begin{equation}
\rho^2 (V_{\rho \eta}^2 + V_{\eta \eta}^2) \to \rho^2 f(\eta)^2, \qquad w \to \frac{1}{\rho^2} \frac{g(\eta)}{f(\eta)^2}, \qquad A \to - \frac{h(\eta)}{f(\eta)^2} \, \dd \phi,
\end{equation}
where the functions $f(\eta), g(\eta), h(\eta)$ are given by
\begin{align}
f(\eta) &= \sum_{i=1}^N \frac{k^3_i}{ \abs{\eta - \eta_i} }, \\
g(\eta) &= \bigg( q_0 - \sum_{i=1}^N q_i \sign (\eta - \eta_i) \bigg) f(\eta), \\
\begin{split}
h(\eta) &= \sum_{i=1}^N \frac{q_i}{\abs{\eta - \eta_i}} \, f(\eta) \\
& \qquad \qquad + \frac12 \bigg( q_0 - \sum_{i=1}^N q_i \sign (\eta - \eta_i) \bigg) \sum_{j=1}^N \frac{k^3_j \sign (\eta - \eta_j)}{(\eta - \eta_j)^2}.
\end{split}
\end{align}
Then as $\rho \to 0$, the 4-metric can be rearranged to give
\begin{equation}
g \to \frac{g(\eta)}{f(\eta)^2} \, \dd \phi + \frac{f(\eta)^2}{g(\eta)} \bigg[ \frac{g(\eta)^2}{f(\eta)^2} ( \dd \rho^2 + \dd \eta^2 ) + \rho^2 \, \dd \tau^2 \bigg],
\end{equation}
where the coordinates $\tau, \phi$ have now exchanged roles.  Notably, along the entire segment over which $a$ (defined in \eqref{a def}) vanishes, the $\phi$ circle remains a finite size as $\rho \to 0$, whereas the $\tau$ circle pinches off.  In particular, we have
\begin{equation}
\frac{g(\eta)^2}{f(\eta)^2} = \bigg( q_0 - \sum_{i=1}^N q_i \sign (\eta - \eta_i) \bigg)^2 \equiv 4 b^2,
\label{tau pinch}
\end{equation}
so the $\tau$ circle is pinching off in a conical singularity $\RR^2 / \ZZ_b$ (the factor of 4 in \cref{tau pinch} is to account for the fact that the period of $\tau$ is $4 \pi$ rather than $2 \pi$).  This sort of homology 2-cycle, in which $\phi$ remains finite while $\tau$ pinches off along a finite portion of the axis, is illustrated in  \cref{phi cycles}.

We also point out that the axisymmetric LeBrun metrics we consider here are toric K\"ahler manifolds, and there is possibly a more elegant description of what is going on with the various types of 2-cycles using the techniques of toric geometry \cite{Abreu2009}.

\subsection{Magnetic flux through cycles}
\label{base flux}

A desired property of these new solutions is that the magnetic 2-form $\Th 3$ have non-trivial flux through the homological 2-cycles in the base.  The 2-form $\Th 3$ is given by
\begin{equation} \label{Th full}
\Th 3 = \frac12 (\dd \tau + A) \wedge \dd \frac{u_z}{w} + \frac12 w \hodge_3 \dd \frac{u_z}{w},
\end{equation}
but it will be more helpful to write it as
\begin{equation} \label{B3}
\Th 3 = \dd B^3 = - \frac 12 \dd \Big[ \frac{u_z}{w} (\dd \tau + A) +  r u_r \, \dd \phi \Big]
\end{equation}
where
\begin{gather}
\frac12 u_z =  \frac{V_{\eta \eta}}{\rho^2 (V_{\rho \eta}^2 + V_{\eta \eta}^2)}, \qquad \frac12 r u_r = -1 + \frac{1}{\rho (V_{\rho \eta}^2 + V_{\eta \eta}^2)} V_{\rho \eta}, \\
w =  \frac{1}{\rho (V_{\rho \eta}^2 + V_{\eta \eta}^2)} \big( V_{\eta \eta} \, \hat w_\rho - V_{\rho \eta} \, \hat w_\eta \big).
\end{gather}
On a 2-cycle $\Delta_{ij}$ swept out by the $\tau$ fiber, the flux can be computed via
\begin{equation} \label{flux 3}
\Flux{3}_{ij} = \frac{1}{4\pi} \int_{\Delta_{ij}} \Th{3} = \frac{1}{4\pi} \int_{\Delta_{ij}} \dd \tau \wedge \dd \frac{K^3}{w} = \frac{k_j}{\widetilde{q}_j} - \frac{k_i}{\widetilde{q}_i},
\end{equation}
where $\widetilde q_i \equiv q_i (\bar K^3_i - k^3_0) - k^3_i (\bar Q_i - q_0)$.  This is very reminiscent of the fluxes in the BPS case \cite{Bena:2007kg}, and in stark contrast to previous non-BPS work \cite{Bobev:2011kk, Bobev:2012af} where $\Th 3$ had no topological fluxes.

On a 2-cycle formed by the $\phi$ circle, one has to be considerably more careful.  Along a line segment of the $\eta$ axis between $\eta_i$ and $\eta_j$ where the $\phi$ circle has a finite size, one can show that as $\rho \to 0$,
\begin{equation} \label{Th phi cycle}
\Th 3 \to \frac{1}{g_0} \, \dd \Big[ - \dd \tau + \frac{\tilde f(\eta)}{f(\eta)} \, \dd \phi \Big],
\end{equation}
where
\begin{equation}
f(\eta) = \sum_{i=1}^N \frac{k^3_i}{ \abs{\eta - \eta_i} }, \qquad \tilde f(\eta) = \sum_{i=1}^N \frac{q_i}{ \abs{\eta - \eta_i} }, \qquad g_0 = \bigg( q_0 - \sum_{i=1}^N q_i \sign (\eta - \eta_i) \bigg),
\end{equation}
and we note that along this single line segment between two GH points, $g_0$ is constant.  {\it Outside} this line segment, the approximation \eqref{Th phi cycle} no longer holds; in particular, we should not be concerned about the $\sign (\eta - \eta_i)$ in $g_0$, because the full $\Th 3$ \eqref{Th full} is continuous everywhere and has no jumps.  Then using \eqref{Th phi cycle}, the flux of $\Th 3$ through a $\phi$ cycle is given by
\begin{equation}
\Flux{3}_{ij} = \frac{1}{4 \pi} \int_{\Delta_{ij}} \frac{1}{g_0} \, \dd \frac{\tilde f(\eta)}{f(\eta)} \wedge \dd \phi = \frac{1}{2 g_0} \bigg( \frac{q_j}{k^3_j} - \frac{q_i}{k^3_i} \bigg),
\end{equation}
which, interestingly, has a very different structure to \eqref{flux 3}.

Therefore we have succeeded in constructing a useful base space.  It has the homological 2-spheres we expected, swept out by $\tau$; these have cohomological fluxes which can be adjusted in any desired way by choosing parameters.  As a bonus, we also obtain homological 2-spheres swept out by $\phi$, which also have cohomological flux.

Interestingly, the fluxes of each type take different forms.  If we assign units to the parameters of the solution, then $\tau$ fluxes have units of ``$1/q$" and $\phi$ fluxes have units of ``$1/k$".  This is consistent with the coordinate transformation \eqref{tau trans}, \eqref{phi trans}; if we assume the angles $\psi, \chi$ are dimensionless, then the the fluxes $\Flux{3}_{ij}$ will have the same units through both $\tau$ cycles and $\phi$ cycles.

\section{Multi-centered supergravity solutions}

Now that we have an appropriate base space, we must solve the system \eqref{L1 L2 eqns}, \eqref{K1 eqn}, \eqref{K2 eqn}, \eqref{M eqn}, \eqref{L3 eqn}, and finally \eqref{omega eqn}.  The route to the solutions is tedious and not particularly illuminating, so we refer the reader to Appendix \ref{solutions} for the details, including the full, explicit solutions themselves.  In this section, we will focus on analyzing the solutions.

The solutions are described by $N$ number of points $\eta_i$ along the axis in the base space, and by the $8N + 10$ parameters $\{q_0, k^1_0, k^2_0, k^3_0, \ell_1^0, \ell_2^0, \ell_3^0, m_0, \omega_0, \ell_3^z, q_i, k^1_i, k^2_i, k^3_i, \ell_1^i, \ell_2^i, \ell_3^i, m_i\}$.  The following sections make frequent reference to these parameters as they appear in the solutions of Appendix \ref{solutions}.

\subsection{Asymptotics of the 5d metric}
\label{asymptotics}

We should first look at the behavior of the 5-dimensional metric \eqref{5d metric} at infinity.  We leave the details to Sec. \ref{asymptotics app}, and summarize the main results here.

We define the coordinates $R, \theta$ via
\begin{equation}
\rho = R \sin \theta, \qquad \eta = R \cos \theta,
\end{equation}
and look at the various metric functions as $R \to \infty$.  First, we find that the warp factors $Z_1, Z_2$ go as $1/R$:
\begin{equation}
Z_1 \sim \frac{1}{R}, \qquad Z_2 \sim \frac{1}{R}.
\end{equation}
The functions $\mu, \omega_{(\phi)} \sim (\text{const})$ at infinity, but to avoid CTC's, we must choose parameters such that these constants vanish.  At the $1/R$ order, these functions pick up an angular dependence\footnote{The reason for labelling these functions ``5, 6'' will become apparent in the next subsection.} on $\theta$:
\begin{equation}
\mu \sim \frac{1}{R} f_5(\theta), \qquad \omega_{(\phi)} \sim \frac{1}{R} f_6(\theta).
\end{equation}

Next one is interested in $Z_3$, and one has a choice.  The leading order behavior is constant:
\begin{equation}
Z_3 \sim \ell_3^0 - \sum_{\substack{ij \\ i \neq j}} \frac{k^1_i \ell_1^j + k^2_i \ell_2^j - k^3_i \ell_3^j + 2 m_i q_j}{\eta_i - \eta_j} + \mathcal{O} \Big( \frac{1}{R} \Big). \label{Z3 first}
\end{equation}
However, as mentioned in Sec. \ref{floating branes}, the $Z_I$ must all have the same asymptotic behavior to allow an M-theory lift.  Hence we should choose $\ell_3^0$ to make the constant term vanish in \eqref{Z3 first}.  Alternatively, one can keep the constant term, allowing $Z_3$ to have different behavior to $Z_1, Z_2$---as was pointed out in \cite{Bobev:2012af}, this can be lifted naturally to the 6-dimensional theory obtained by reducing IIB supergravity on $T^4$.

\subsubsection{Asymptotics for lifting to 11d SUGRA}

We first consider the case that all three $Z_I$ have the same asymptotic behavior.  Therefore the leading order constant $Z_3$ \eqref{Z3 1st} must vanish, hence we set: 
\begin{equation}
\ell_3^0 = \sum_{\substack{ij \\ i \neq j}} \frac{k^1_i \ell_1^j + k^2_i \ell_2^j - k^3_i \ell_3^j + 2 m_i q_j}{\eta_i - \eta_j}. \label{l30 const}
\end{equation}
The 5-dimensional metric \eqref{5d metric} then becomes
\begin{equation} \label{5d asym form}
\begin{split}
\dd s_5^2 &= - \frac{R^2}{f_4(\theta)^2} \Big[ \dd t + \frac{1}{R} f_5(\theta) \, \dd \tau + \frac{1}{R} \Big( f_5(\theta) f_3(\theta) + f_6(\theta) \Big) \, \dd \phi \Big]^2 \\
& \qquad \qquad + \frac{f_4(\theta)}{f_2(\theta)} \Big( \dd \tau + f_3(\theta) \, \dd \phi \Big)^2 \\
& \qquad \qquad \qquad + \frac{f_2(\theta) f_4(\theta)}{R^2} \Big[ f_1(\theta) (\dd R^2 + R^2 \, \dd \theta^2) + R^2 \sin^2 \theta \, \dd \phi^2 \Big],
\end{split}
\end{equation}
where generically speaking,
\begin{gather}
\rho^2 (V_{\rho \eta}^2 + V_{\eta \eta}^2 ) \sim f_1(\theta), \qquad w \sim \frac{1}{R} f_2(\theta), \qquad A \sim f_3(\theta) \, \dd \phi \\
Z \sim \frac{1}{R} f_4(\theta), \qquad \mu \sim \frac{1}{R} f_5(\theta), \qquad \omega \sim \frac{1}{R} f_6(\theta) \, \dd \phi,
\end{gather}
and simplifications likely occur in \cref{5d asym form} if one works these out in more specificity.  Due to the $\dd R^2 / R^2$ term, this metric is something related to $AdS_2 \times S^3$.  Specifically, it is a warped, rotating quotient $AdS_2 \times S^3 / G_\infty$, where $G_\infty$ is a finite group acting on the $S^3$ factor as described in \cref{base asym}.

If we choose parameters such that $\widetilde q_\infty = \pm 1$ as defined in \cref{q infty def}, then the base space approaches $\RR^4$ without orbifold identifications, as described in \cref{ambipolar}.  One can then choose parameters such that
\begin{equation}
Z_3 \sim \frac{1}{R}, \qquad \mu \sim \frac{1}{R} (c_1 + c_2 \cos \theta), \qquad \omega \sim \cO(R^{-2}),
\end{equation}
and therefore $Z \equiv (Z_1 Z_2 Z_3)^{1/3} \sim 1/R$, without angular dependence.  Then changing coordinates via
\begin{equation}
R = \frac14 \varrho^2, \qquad \theta = 2 \vartheta, \qquad \tau = \psi + \chi, \qquad \phi = \psi - \chi,
\end{equation}
(up to shifts in $t$ and $\tau$), one obtains a 5-dimensional metric of the form
\begin{equation} \label{bmpv}
\dd s_5^2 = - \varrho^4 \Big( \dd t + J_1 \frac{\sin^2 \vartheta}{\varrho^2} \, \dd \psi + J_2 \frac{\cos^2 \vartheta}{\varrho^2} \, \dd \chi \Big)^2 + \frac{\dd \varrho^2}{\varrho^2} + \dd \Omega_3^2,
\end{equation}
which is the metric of the near-horizon region of a BMPV black hole \cite{Breckenridge:1996is}.

\subsubsection{Asymptotics lifting to IIB on $T^4$}

Alternatively, we can choose to allow $Z_3 \sim (\text{const})$ at infinity while $Z_1, Z_2 \sim 1 / \varrho^2$, and therefore not impose \eqref{l30 const}.  Then the 5-dimensional metric will generically be of the form
\begin{equation}
\dd s_5^2 = - \varrho^{8/3} \, (\dd t + k)^2 + \varrho^{-4/3} \, (\dd \varrho^2 + \varrho^2 \, \dd \Omega_3^2),
\end{equation}
which looks somewhat strange.  As shown in \cite{Bobev:2012af}, however, there is a natural lift into 6-dimensional $\Neql 1$ supergravity coupled to one anti-self-dual tensor multiplet \cite{Gutowski:2003rg, Cariglia:2004kk, Bena:2011dd}.  The metric ansatz in 6 dimensions can be written in terms of the 5-dimensional quantities as
\begin{equation}
\dd s_6^2 = - \frac{2}{\sqrt{Z_1 Z_2}} \, \big( \dd v + B^3 \big) \bigg( \dd u + k - \frac12 Z_3 \, \big( \dd v + B^3 \big) \bigg) + \sqrt{Z_1 Z_2} \, \dd s_4^2,
\end{equation}
where $B^3$ is the 1-form potential such at $\Th 3 = \dd B^3$ as in \eqref{B3}.  In this context, applying the asymptotics at infinity where $Z_3 \sim (\text{const})$ and $Z_1, Z_2 \sim 1 / \varrho^2$ gives the result
\begin{equation} \label{pp wave}
\dd s_6^2 = - 2 \varrho^2 \, \dd v \, \big( \dd u + k - \frac12 Z_3 \, \dd v) + \frac{\dd \varrho^2}{\varrho^2} + \dd \Omega_3^2,
\end{equation}
which is a momentum wave propagating on $AdS_3 \times (S^3 / G_\infty)$.  Furthermore, nothing prevents us from imposing $Z_3 \sim 1/\varrho^2$ in this lifted metric; in such a case, one would obtain the 6-dimensional lift of the near-horizon BMPV metric \eqref{bmpv}, which is the near-horizon metric of a BPS, rotating D1-D5-P black string \cite{Giusto:2004id}.

\subsubsection*{Summarizing asymptotics}

Generally speaking, we see that our solutions are asymptotic to a warped, rotating version of $AdS_2 \times (S^3 / G_\infty)$, and for special choices of parameters, to near-horizon BMPV.  Alternatively, one can lift to IIB supergravity on $T^4$, giving a 6-dimensional metric which allows $Z_3$ to have different asymptotics to $Z_1, Z_2$.  In this case, one can impose $Z_3 \sim (\text{const})$ to obtain a momentum wave solution propagating on $AdS_3 \times (S^3 / G_\infty)$; or, imposing $Z_3 \sim 1/\varrho^2$, one obtains the near-horizon metric of a BPS, rotating black string.

We should note from constraints derived in \cite{Bobev:2011kk}, that the ``floating brane" equations \cite{Bena:2009fi} on a K\"ahler base do not have asymptotically flat solutions, and solutions must generically have nonzero rotation parameters at infinity.  The reason for this is that the $T_{00}$ component of the 5-dimensional energy-momentum tensor is a manifestly positive-definite function of the $Z_I, \Th I$.  If we have $Z_I \sim 1$ at infinity, then $\Th 1, \Th 2$ still contain a term proportional to the K\"ahler form $J$, which contributes a constant to $T_{00}$ and prevents asymptotic flatness.  The rotation at infinity comes from the off-diagonal terms $T_{0a}$, which also do not vanish.

\subsection{Regularity conditions}

The solutions we have obtained generically have a number of singularities at each $\eta_i$ which act as sources of the electric potentials $Z_I$ and magnetic field strengths $\Th I$.  However, in the context of black hole microstate geometries, we are interested in solutions that are everywhere smooth, with no singular sources.  This can be accomplished by choosing the parameters in such a way that singularities are eliminated.  The necessary condition for smoothness is that each of the functions $Z_1, Z_2, Z_3, \mu, \omega_{(\phi)}$ remain non-singular as the GH points are approached.

Looking near a point $\eta_\ell$, we again define a local radial coordinate via
\begin{equation}
\rho = R \sin \theta, \qquad \eta - \eta_\ell = R \cos \theta.
\end{equation}
Then as $R \to 0$, we have
\begin{align}
Z_1 &\to \frac{1}{R} \bigg( \frac{k^2_\ell k^3_\ell + q_\ell \ell_1^\ell}{q_\ell \big( \bar K^3_\ell - k^3_0 \big) - k^3_\ell \big( \bar Q_\ell - q_0 \big)} \bigg), \\
Z_2 &\to \frac{1}{R} \bigg( \frac{k^1_\ell k^3_\ell + q_\ell \ell_2^\ell}{q_\ell \big( \bar K^3_\ell - k^3_0 \big) - k^3_\ell \big( \bar Q_\ell - q_0 \big)} \bigg),
\end{align}
where again,
\begin{equation}
\bar K^3_\ell \equiv \sum_{\substack{i \\ i \neq \ell}} k^3_i \sign (\eta_\ell - \eta_i), \qquad \bar Q_\ell \equiv \sum_{\substack{i \\ i \neq \ell}} q_i \sign (\eta_\ell - \eta_i).
\end{equation}
Therefore, the singular parts of $Z_1, Z_2$ will vanish if
\begin{equation} \label{l1 l2 reg}
\ell_1^\ell = - \frac{k^2_\ell k^3_\ell}{q_\ell}, \qquad \ell_2^\ell = - \frac{k^1_\ell k^3_\ell}{q_\ell},
\end{equation}
at every GH point.  Next, imposing \eqref{l1 l2 reg}, we have
\begin{equation}
\begin{split}
Z_3 &\to \frac{1}{R} \bigg[ \frac{k^1_\ell k^2_\ell}{q_\ell^2} \Big( q_\ell (\bar K^3_\ell - k^3_0) - k^3_\ell (\bar Q_\ell - q_0) \Big) - \ell_3^\ell (\bar K^3_\ell - k^3_0) + 2 m_\ell (\bar Q_\ell - q_0) \\
& \qquad \qquad - \Big( k^3_\ell \ell_3^\ell + 2 m_\ell q_\ell \Big) \cos \theta \bigg],
\end{split}
\end{equation}
and hence the singular part of $Z_3$ vanishes if
\begin{equation} \label{l3 m reg}
\ell_3^\ell = \frac{k^1_\ell k^2_\ell}{q_\ell}, \qquad m_\ell = - \frac{k^1_\ell k^2_\ell k^3_\ell}{2 q_\ell^2}.
\end{equation}
Together, \eqref{l1 l2 reg} and \eqref{l3 m reg} are also sufficient to guarantee $\mu \to (\text{const})$ and $\omega_{(\phi)} \to (\text{const})$ near $\eta_\ell$; hence we will have a regular solution if we impose these conditions at every GH point.

We note that these conditions appear exactly the same (up to signs that result from differing conventions) as those in the original BPS story \cite{Bena:2007kg}.  However, there is a key difference:  In these solutions, the parameters $q_\ell$ do not directly control the singularities of $w$, but as in \eqref{near points}, the singularities in $w$ are controlled by the determinants
\begin{equation}
\widetilde{q}_\ell \equiv q_\ell (\bar K^3_\ell - k^3_0) - k^3_\ell (\bar Q_\ell - q_0).
\end{equation}

\subsection{Fluxes through cycles}
\label{phys flux}

It will be useful to have expressions for the magnetic flux threading 2-cycles formed by sweeping the $\tau$ fiber between GH points in the 4-dimensional base space.  We have already calculated the flux of $\Th 3$ on these cycles \eqref{flux 3}:
\begin{equation}
\Flux{3}_{ij} \equiv \frac{1}{4\pi} \int_{\Delta_{ij}} \Th{3}= \frac{k_j}{\widetilde{q}_j} - \frac{k_i}{\widetilde{q}_i}.
\end{equation}
Before computing the remaining two fluxes, we will impose the regularity conditions \eqref{l1 l2 reg}, \eqref{l3 m reg}.  Then as we approach a GH point $\eta_\ell$, we have
\begin{equation}
\frac{K^1}{w} \to \frac{k^1_\ell (\bar K^3_\ell - k^3_0)}{q_\ell} - \ell_2^0 + \bar L_2^\ell, \qquad \frac{K^2}{w} \to \frac{k^2_\ell \bar (K^3_\ell - k^3_0)}{q_\ell} - \ell_1^0 + \bar L_1^\ell,
\end{equation}
where we have defined new quantities
\begin{equation}
\bar L_1^\ell \equiv \sum_{\substack{i \\ i \neq \ell}} \ell_1^i \sign (\eta_\ell - \eta_i), \qquad \bar L_2^\ell \equiv \sum_{\substack{i \\ i \neq \ell}} \ell_2^i \sign (\eta_\ell - \eta_i).
\end{equation}
Then the flux through $\tau$ cycles can be computed in a way similar to \eqref{flux 3}:
\begin{align}
\flux{1}_{ij} &\equiv \frac{1}{4\pi} \int_{\Delta_{ij}} \Th{1} = \frac{k^1_j (\bar K^3_j - k^3_0)}{q_j} + \bar L_2^j - \frac{k^1_i (\bar K^3_i - k^3_0)}{q_i} - \bar L_2^i, \label{flux 1} \\
\flux{2}_{ij} &\equiv \frac{1}{4\pi} \int_{\Delta_{ij}} \Th{2} = \frac{k^2_j (\bar K^3_j - k^3_0)}{q_j} + \bar L_1^j - \frac{k^2_i (\bar K^3_i - k^3_0)}{q_i} - \bar L_1^i. \label{flux 2}
\end{align}
One can in principle also compute the fluxes through the 2-cycles swept out by $\phi$, as was done in Sec. \ref{base flux}.  However, this is tedious and of no special benefit to the rest of this analysis, so we omit it.

\subsection{Causality conditions:  the ``bubble equations"}

We have determined the conditions that a solution is smooth as one approaches the various Gibbons-Hawking points in the base manifold.  However, to construct sensible supergravity solutions, one must also ensure that there are no closed timelike curves.

Looking at the metric \eqref{5d metric} on a surface of constant $t$, we can rearrange it as follows:
\begin{equation} \label{ctc metric}
\begin{split}
\dd s_5^2 &= \frac{\cQ}{w^2 Z^2} \bigg( \dd \tau + A - \frac{w^2 \mu}{\cQ} \, \omega \bigg)^2 + Z w \bigg( \rho^2 \, \dd \phi^2 - \frac{\omega^2}{\cQ} \bigg) \\
& \qquad \qquad + Z w \, \rho^2 (V_{\rho \eta}^2 + V_{\eta \eta}^2) (\dd \rho^2 + \dd \eta^2),
\end{split}
\end{equation}
where
\begin{equation}
\cQ \equiv Z_1 Z_2 Z_3 w - w^2 \mu^2, \qquad Z \equiv (Z_1 Z_2 Z_3)^{1/3}.
\end{equation}
In order for CTC's to be absent everywhere, \eqref{ctc metric} must be positive-definite.  This requires
\begin{equation}
\cQ \geq 0, \qquad Z w \geq 0, \qquad \rho^2 \, \dd \phi^2 \geq \frac{\omega^2}{\cQ}.
\end{equation}
It is generally impractical to enforce these global conditions from the local point of view of choosing parameters in the solution; one must write down a solution and then explore it numerically to look for CTC's.  However, one can look at \emph{local} causality conditions near the GH points, and this leads to a system of equations that must be solved as a necessary (but not sufficient) condition that a solution be causally sensible.

In the BPS context \cite{Bena:2007kg}, this leads to a system of so-called ``bubble equations" that relate the distances between the GH centers (as measured in the $\RR^3$) to the product of the fluxes of the $\Th I$ through the various 2-cycles described by the GH centers.  Thus the size of each ``bubble" is governed by the amount of flux trapped on it.  Importantly, the bubble equations depend upon the product of all three fluxes.  In previous work on non-supersymmetric solutions derived from floating branes \cite{Bobev:2011kk, Bobev:2012af}, the third flux $\Th 3$ was topologically trivial and contributed no fluxes to the bubble equations.  The result was that the causality conditions did not constrain the sizes of the homological 2-cycles.  In these new solutions, however, $\Th 3$ has non-trivial fluxes on the 2-cycles (as in Sec. \ref{base flux}), so we expect to find non-trivial bubble equations.

Looking at \eqref{ctc metric} near the GH points, one finds two potential sources of CTC's coming from the two angular coordinates $\tau, \phi$.  To eliminate CTC's near the GH points, we must require that
\begin{equation}
\mu \to 0, \qquad \omega \to 0
\end{equation}
at these points.  While these appear to be two different conditions, they are really the same.  To see this, we can rearrange the $\omega$ equation \eqref{omega eqn} as follows:
\begin{equation} \label{omega eqn 2}
\begin{split}
\dd \omega &= w Z_1 \hodge_3 \dd \frac{K^1}{w} + w Z_2 \hodge_3 \dd \frac{K^2}{w} + w Z_3 \hodge_3 \dd \frac{K^3}{w} - 2 w Z_1 Z_2 \hodge_3 \dd z \\
& \qquad + w \hodge_3 \dd \mu - \mu \, \dd A.
\end{split}
\end{equation}
We choose parameters such that $\omega$ vanishes at infinity (as in \eqref{m0 soln}, \eqref{w0 soln}), so for $\omega$ to be non-vanishing somewhere on the axis would require Dirac-Misner strings.  Given the regularity conditions \eqref{l1 l2 reg}, \eqref{l3 m reg}, the only term in \eqref{omega eqn 2} that can source Dirac-Misner strings is $ - \mu \, \dd A$.  Therefore, to eliminate local CTC's near the GH points, it is enough to demand that $\mu$ vanish at each GH point.  The vanishing of $\mu$ results in the following ``bubble equations" at each $\eta_\ell$:
\begin{equation} \label{bubble}
\begin{split}
- 2 m_0 \widetilde q_\ell + k^3_0 \ell_3^0 &~=~ (k^3_0 - \bar K^3_\ell) \sum_{\substack{i \\ i \neq \ell}} \hFlux{1}_{\ell i} \hFlux{2}_{\ell i} \hFlux{3}_{\ell i} \frac{q_\ell q_i}{r_{\ell i}} \\
& \qquad \qquad + \frac12 k^3_\ell \sum_{\substack{ij \\ i \neq j}} \hFlux{1}_{ij} \hFlux{2}_{ij} \hFlux{3}_{ij} \frac{q_i q_j}{r_{ij}} \, s(i,j) \, s(\ell,i) \, s(\ell,j)
\end{split}
\end{equation}
where we have defined
\begin{gather}
r_{ij} \equiv \abs {\eta_i - \eta_j}, \qquad \hflux{I}_{ij} \equiv \bigg( \frac{k^I_j}{q_j} - \frac{k^I_i}{q_i} \bigg), \qquad s(a,b) \equiv \sign (\eta_a - \eta_b), \\
\widetilde{q}_\ell \equiv q_\ell (\bar K^3_\ell - k^3_0) - k^3_\ell (\bar Q_\ell - q_0).
\end{gather}
The combinations of parameters $\hflux{I}_{ij}$ which appear in the bubble equations are not the \emph{physical} fluxes $\flux{I}_{ij}$ calculated in \cref{flux 3,flux 1,flux 2}.  However, with a little algebra one can show that they are related linearly and homogeneously\footnote{Here we again assume the regularity conditions \cref{l1 l2 reg}, \cref{l3 m reg} are imposed.}:
\begin{align}
\flux{1}_{\ell i} &= (- k^3_0 + \bar K^3_\ell ) \, \hflux{1}_{\ell i} + \sum_{j = 1}^N k^3_j \, \hflux{1}_{ij} \, \big( s(\ell, j) - s(i,j) \big), \label{hflux 1} \\
\flux{2}_{\ell i} &= (- k^3_0 + \bar K^3_\ell ) \, \hflux{2}_{\ell i} + \sum_{j = 1}^N k^3_j \, \hflux{2}_{ij} \, \big( s(\ell, j) - s(i,j) \big), \label{hflux 2} \\
\widetilde q_\ell \widetilde q_i \, \flux{3}_{\ell i} &=  q_\ell q_i (- k^3_0 + \bar K^3_\ell ) \, \hflux{3}_{\ell i} + k^3_\ell \sum_{j=1}^N q_i q_j \, \hflux{3}_{ij} \big( s(\ell, j) - s(i,j) \big). \label{hflux 3}
\end{align}
These look tantalizingly like they might allow a simpler expression of the right-hand side of \cref{bubble}; however, the presence of $1/r_{\ell i}, 1/r_{ij}$ in the sums complicates the algebra, and the expression we have written in \cref{bubble} is probably the simplest.

We have thus succeeded in finding a family of non-BPS solutions with \emph{non-trivial} bubble equations which constrain the bubble diameters $r_{ij}$ in terms of the fluxes trapped on the bubbles.  We also observe that there is a significant, important difference between these non-BPS bubble equations and the well-known BPS version \cite{Bena:2007kg}.  The term on the second line of \cref{bubble} is entirely new:  In order to avoid CTC's at $\eta_\ell$, the equations depend not only on the diameters $r_{\ell i}$ of the 2-cycles adjacent to $\eta_\ell$, but also on the diameters $r_{ij}$ of each of the other 2-cycles.  This is telling us about \emph{new physics}:  these \emph{non}-supersymmetric solutions exhibit a richer variety of $E \times B$ interactions than previously known BPS solutions.

However, while these bubble equations differ from the BPS ones in a few ways, they are similar in a particularly striking way:  They are \emph{linear} in the inverse bubble diameters $1/r_{ij}$.  This stands in contrast to the so-called ``almost BPS" family of solutions where the bubble equations are cubic in the inverse distances \cite{Bena:2009en, Bena:2013gma, Vasilakis:2011ki}.  So although these solutions lack supersymmetry, they are in some sense closer to BPS than the ``almost BPS" solutions.  This is because they are trivial KK reductions of 6-dimensional geometries which are BPS in the IIB frame \cite{Bobev:2012af}.

\subsubsection*{Number of independent bubble equations}

Ultimately, there are only $N-1$ independent $r_{ij}$, so we expect there to be $N-1$ independent bubble equations.  This is easiest to demonstrate by looking directly at the Dirac-Misner strings in $\omega$.  This results in the same bubble equations as above, each multiplied by a constant (which is different at each $\eta_\ell$).  Near $\eta_\ell$, the Dirac-Misner string part of $\omega$ is given by the jump that occurs in crossing from one side of $\eta_\ell$ to the other:
\begin{equation} \label{bubble coefs}
\omega \Big\rvert_{\theta = 0} - \omega \Big\rvert_{\theta = \pi} = - \bigg( A \Big\rvert_{\theta = 0} - A \Big\rvert_{\theta = \pi} \bigg) \mu = \frac{2 \, \widetilde q_\ell}{\big( \bar K^3_\ell - k^3_0 \big)^2 - \big( k^3_\ell \big)^2} \, \mu \; \dd \phi.
\end{equation}
Since $\omega$ contains a sequence of Dirac-Misner string sources along the $\eta$ axis, and vanishes at both positive and negative infinity, then the sum of all the jumps must be zero.  Therefore, the weighted sum of all the bubble equations \eqref{bubble}, each multiplied by the coefficient in \eqref{bubble coefs}, must give zero.  This weighted sum gives
\begin{equation}
m_0 = \frac12 \frac{K^3_\star}{q_0 K^3_\star - k^3_0 Q_\star} \bigg( \ell_3^0 - \sum_{\substack{ij \\ i \neq j}} \frac{k^1_i \ell_1^j + k^2_i \ell_2^j - k^3_i \ell_3^j + 2 m_i q_j}{\eta_i - \eta_j} \bigg).
\end{equation}
which is the condition we have already imposed \eqref{m0 soln} in order that $\mu \to 0$ at infinity.  Hence as expected, the bubble equations constitute $N-1$ independent equations in the $N-1$ independent variables $r_{ij}$.

\subsubsection*{Hints of scaling solutions}

Finally, there is a curious thing that happens if we impose all of the conditions derived in \cref{asymptotics} for near-horizon BMPV-like (i.e. warped, rotating $AdS_2 \times S^3$) asymptotics.  First we note that the value of $\ell_3^0$ in \cref{l30 const} is entirely a linear combation of the inverse bubble diameters $1/r_{ij}$.  Second, when \cref{l30 const} is imposed, then $m_0 = \omega_0 = 0$ as in \cref{m0 soln}, \cref{w0 soln}.  Therefore if we insist on near-horizon BMPV-like asymptotics, the bubble equations will take the form, schematically,
\begin{equation}
\sum \hflux{1} \hflux{2} \hflux{3} \frac{q q}{r} = 0.
\end{equation}
If we instead think of this equation as a limiting process where we replace the right-hand side with some $\delta$ and let $\delta \to 0$, then the solutions, as we follow this process, are \emph{scaling solutions} \cite{Bena:2007qc,Bena:2006kb,Vasilakis:2011ki}.  The right-hand side roughly scales as $(\Pi)^3/r$, and thus if we adjust the dipole charges while simultaneously shrinking the bubble diameters, such that $\Pi \sim \lambda, r \sim \lambda$ for $\lambda$ small, this tends toward zero.  In such solutions, the overall size of the bubbled region shrinks (as measured in the 3-dimensional base), while the ratios between the bubble sizes becomes constant.  In the full 5-dimensional metric, this represents the appearance of an arbitrarily deep throat, smoothly capped off by topological bubbles at some finite depth.  Thus one can see the near-horizon BMPV geometry, and the related rotating-$AdS$-like metrics with angular dependence as in \cref{5d asym form}, as the result of this limiting procedure.

More generally, if we consider asymptotic conditions where $Z_3$ behaves differently from $Z_1, Z_2$ (thus naturally lifting to the 6d IIB metric \cref{pp wave} rather than to 11d supergravity), we can set the constant $\ell_3^0$ to anything we like.  In this case, one can find finite, non-trivial solutions to the bubble equations without subjecting them to a limiting procedure.  We demonstrate this in \cref{example sec}.

%

\subsection{An explicit numerical example}
\label{example sec}

In this section we will give an explicit, solved example with three source points, illustrating how a smooth, CTC-free solution can be constructed.  The solution will be in the class asymptotic to \cref{pp wave}, where $Z_3 \sim (\text{const})$ and $Z_1 \sim 1/\rho^2, Z_2 \sim 1/\rho^2$.  We will focus on satisfying the local conditions near the points, and not delve into exactly what asymptotics result.

\begin{figure}
\centering
\includegraphics[width=10cm]{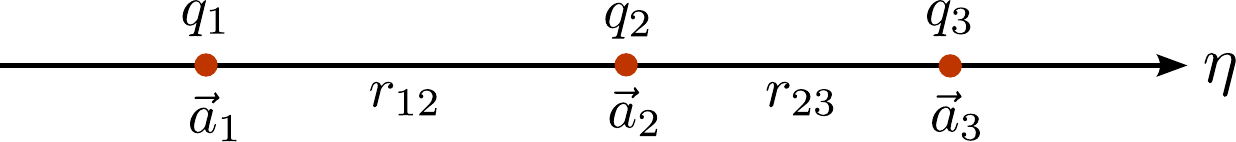}
    \caption{\it \small Setup for a 3-center example.  Geometric charges $q_1, q_2, q_3$ are put at the points $\vec a_1, \vec a_2, \vec a_3$ along the $\eta$ axis.  One must then solve the bubble equations to find $r_{12}, r_{23}$.} 
\label{example}
\end{figure}

We begin by choosing three source points along the $\eta$ axis and assigning them geometric charges.  The parameters of the solution are \emph{ordered} in the manner drawn in \cref{example}; thus by hypothesis the bubble diameters $r_{12}, r_{23}$ are positive.  At the points $\vec a_1, \vec a_2, \vec a_3$ we put the following charges:
\begin{equation} \label{charges}
\begin{aligned}
q_0 &= 2, \quad & \quad q_1 &= 3, \quad & \quad q_2 &= 2, \quad & \quad q_3 &= 6, \\
k^1_0 &= 0, \quad & \quad k^1_1 &= 5, \quad & \quad k^1_2 &= 2, \quad & \quad k^1_3 &= 3, \\
k^2_0 &= 0, \quad & \quad k^2_1 &= 5, \quad & \quad k^2_2 &= 4, \quad & \quad k^2_3 &= 3, \\
k^3_0 &= 1, \quad & \quad k^3_1 &= 2, \quad & \quad k^3_2 &= 2, \quad & \quad k^3_3 &= 2, \\
\ell_1^0 &= 0, \quad & \quad \ell_2^0 &= 0, \quad & \quad \ell_3^0 &= 10, \quad & \quad \ell_3^z &= 0.
\end{aligned}
\end{equation}

Our particular choices are made to satisfy a few constraints:  1) the parity condition \cref{even cond} such that each point will be an orbifold point; 2) the condition that all the $\hflux{I}_{ij}$ are nonzero; 3) the condition that the $\widetilde q_i$ are all ``nice'' numbers; 4) the condition that the bubble equations yield real, positive solutions for the $r_{ij}$; and 5) the condition that $\cQ > 0$ in order to be free of CTC's.  Choosing parameters \cref{charges} to satisfy all of these properties is a bit of an art, and it would be interesting to better understand the moduli space of \emph{physical} solutions.

The value of $\ell_3^0$ sets the overall scale of the solution, as it is the only unconstrained constant sitting on the left-hand side of \cref{bubble}.  Since we have put $\ell_3^0 \neq 0$, this solution will have asymptotics best described in the 6d IIB frame as in \cref{pp wave}.  Most of the functions $w, K^I, L_I, M$ that make up the solution are too lengthy to write out, but as an example, we have
\begin{align}
\hat w_\eta &= \frac{3}{\sqrt{\rho^2 + \eta^2}} + \frac{2}{\sqrt{\rho^2 + (\eta - r_{12})^2}} + \frac{6}{\sqrt{\rho^2 + (\eta - r_{12} - r_{23})^2}}, \\*
\hat w_\rho &= \frac{2}{\rho} - \frac{3 \, \eta}{\rho \, \sqrt{\rho^2 + \eta^2}} - \frac{2 \, (\eta - r_{12})}{\rho \, \sqrt{\rho^2 + (\eta - r_{12})^2}} - \frac{6 \, (\eta - r_{12} - r_{23})}{\rho \, \sqrt{\rho^2 + (\eta - r_{12} - r_{23})^2}},
\end{align}
and so on.  There are two remaining constants $m_0, \omega_0$ which we have not set in \cref{charges}.  To meet the regularity conditions at infinity, these constants will be set equal to \cref{m0 soln,w0 soln}, and then their numerical values will be determined after the $r_{ij}$ are known via solving the bubble equations \cref{bubble}.

At each source point, the base metric approaches $\RR^4 / G_\ell$, where the order of $G_\ell$ at the source point $\eta_\ell$ is given by $\# G_\ell = \abs{\widetilde q_\ell}$, and for the parameters \cref{charges} these $\widetilde q_\ell$ are given by
\begin{equation}
\widetilde q_1 = 5, \qquad \widetilde q_2 = 8, \qquad \widetilde q_3 = 12, \qquad \widetilde q_\infty = 1.
\end{equation}
Therefore we see that this is another example of the phenomenon described in \cref{eng soln}, where the base metric can be asymptotically \emph{globally} flat, despite having orbifold points on the interior, and without resorting to making it ``ambipolar'' as described in \cref{ambipolar}.

\begin{figure}
\centering
\includegraphics[width=13.98cm]{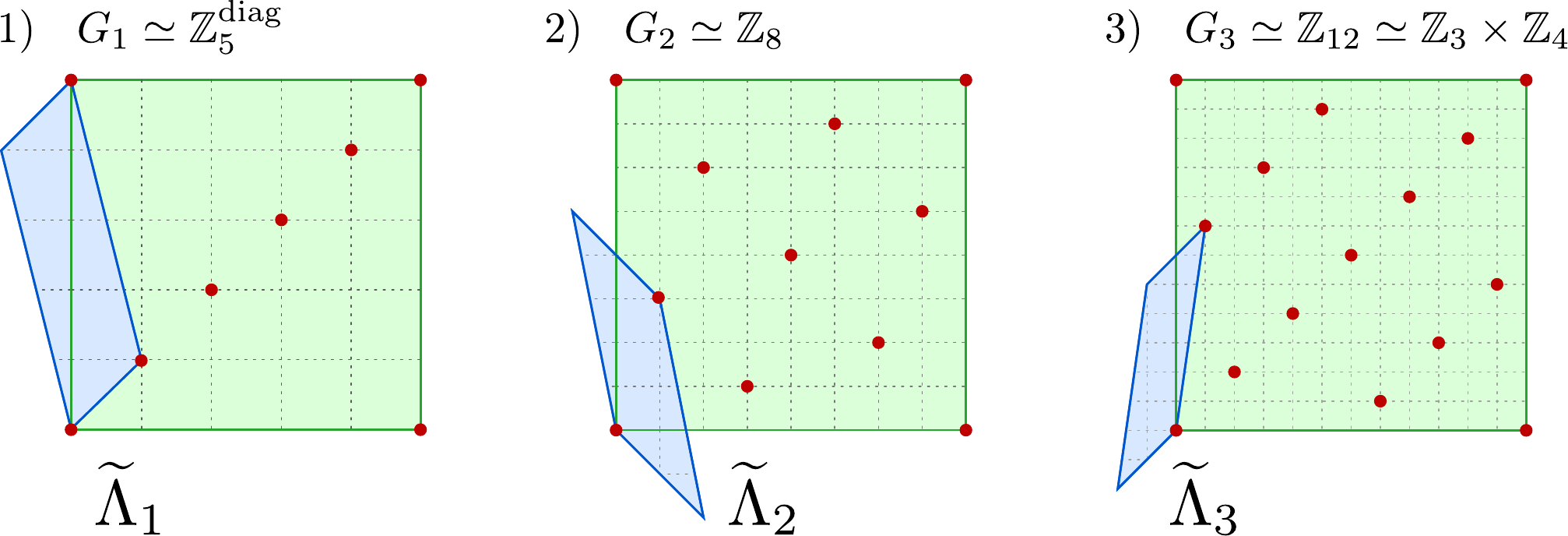}
    \caption{\it \small The unit cells $\widetilde \Lambda_\ell$ of each lattice $\widetilde \Gamma_\ell$ and their corresponding groups $G_\ell \simeq \widetilde \Gamma_\ell / \Gamma$.  The small parallelograms represent the lattice generators \cref{example generators} (where $\widetilde \Lambda_1$ has been shifted by a right $GL(2,\ZZ)$ action in order to make it fit in the figure).  The heavy red dots represent the members of each group $G_\ell$.  The corners of the large squares are to be identified; they represent the lattice $\Gamma$ of the natural $2\pi$ identifications of the $(\alpha, \beta)$ coordinates in $\RR^4$.} 
\label{example quotient}
\end{figure}

We will first analyze the groups at these orbifold points.  We find that the lattice generators $\widetilde \Lambda_\ell$, calculated from \cref{alpha lattice}, are given by
\begin{equation} \label{example generators}
\widetilde \Lambda_1 = \frac15 \begin{pmatrix}2 & -5 \\ -3 & 10\end{pmatrix}, \qquad
\widetilde \Lambda_2 = \frac18 \begin{pmatrix}2 & -1 \\ -2 & 5\end{pmatrix}, \qquad
\widetilde \Lambda_3 = \frac{1}{12} \begin{pmatrix}-2 & 7 \\ -2 & 1\end{pmatrix},
\end{equation}
and the corresponding groups are
\begin{equation}
G_1 \simeq \ZZ_5^\text{diag}, \qquad G_2 \simeq \ZZ_8, \qquad G_3 \simeq \ZZ_{12} \simeq \ZZ_3 \times \ZZ_4,
\end{equation}
where $G_1$ at point $\eta_1$ acts in the diagonal $U(1)$ of $SO(4)$, which one can check using \cref{diag cond}.  These lattice generators $\widetilde \Lambda_\ell$, and the groups given by $G_\ell \simeq \widetilde \Gamma_\ell / \Gamma$, are illustrated in \cref{example quotient}.

Next, we put the general expression for $m_0$ \cref{m0 soln} into the bubble equations \cref{bubble} and solve them for the $r_{ij}$, subject to the triangle constraint
\begin{equation} \label{triangle cond}
r_{12} + r_{23} = r_{13}.
\end{equation}
At this point in the process it is quite possible to fail to find a solution.  The $r_{ij}$ should be strictly positive (they do not enter the equations in a way that allows them to be treated as ``directional'').  The bubble equations are linear in $1/r_{ij}$, and \cref{triangle cond} is linear in $r_{ij}$, hence one is solving a system of quadratic equations.  Thus it is possible to get negative or imaginary $r_{ij}$, and if this happens, one must adjust some of the dipole charges in \cref{charges} and try again.  For the particular charges used here, we obtain two solution sets of real, positive $r_{ij}$, from which we select (via hindsight) the following:
\begin{equation}
r_{12} = 2.45827, \qquad r_{23} = 0.891937, \qquad r_{13} = 3.35021.
\end{equation}
From this solution and the expressions \cref{m0 soln,w0 soln}, we then find
\begin{equation}
m_0 = 1.96384, \qquad \omega_0 = -3.60037,
\end{equation}
which will then guarantee that there are no CTC's at infinity.

\begin{figure}
\centering
\begin{subfigure}{0.3\textwidth}
\centering
\includegraphics[width=\textwidth]{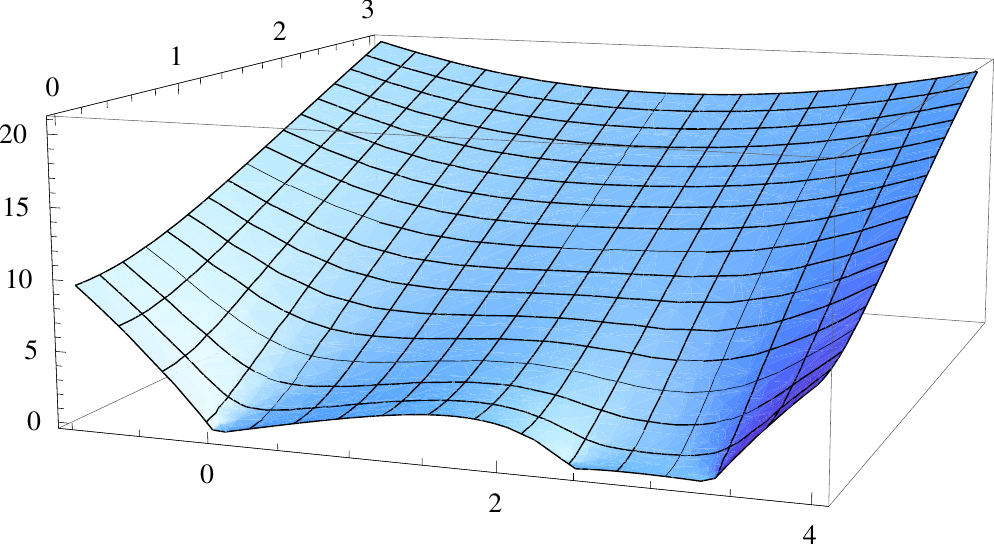}
\end{subfigure}
~
\begin{subfigure}{0.3\textwidth}
\centering
\includegraphics[width=\textwidth]{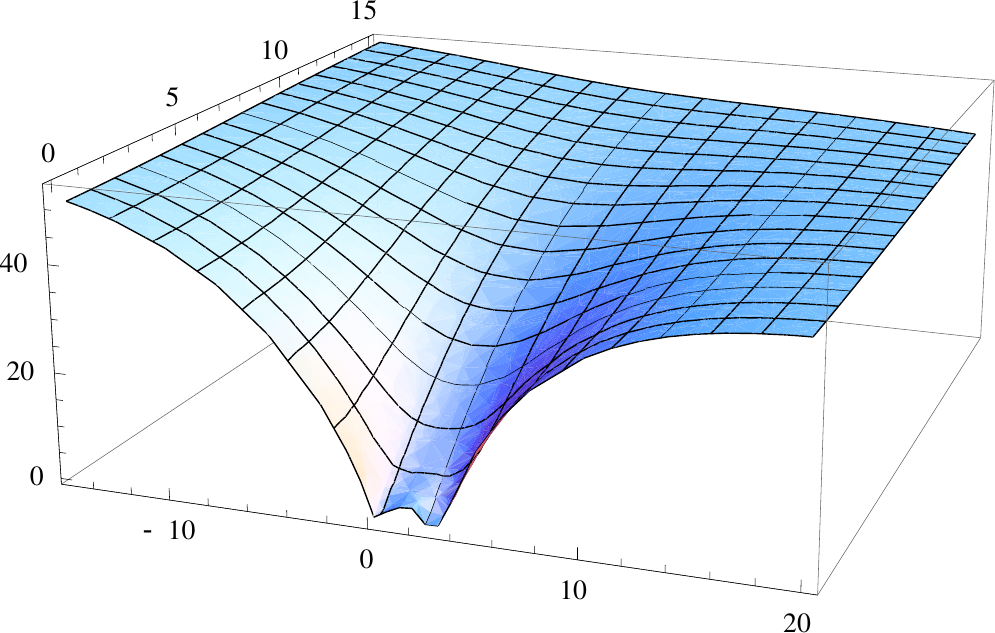}
\end{subfigure}
~
\begin{subfigure}{0.3\textwidth}
\centering
\includegraphics[width=\textwidth]{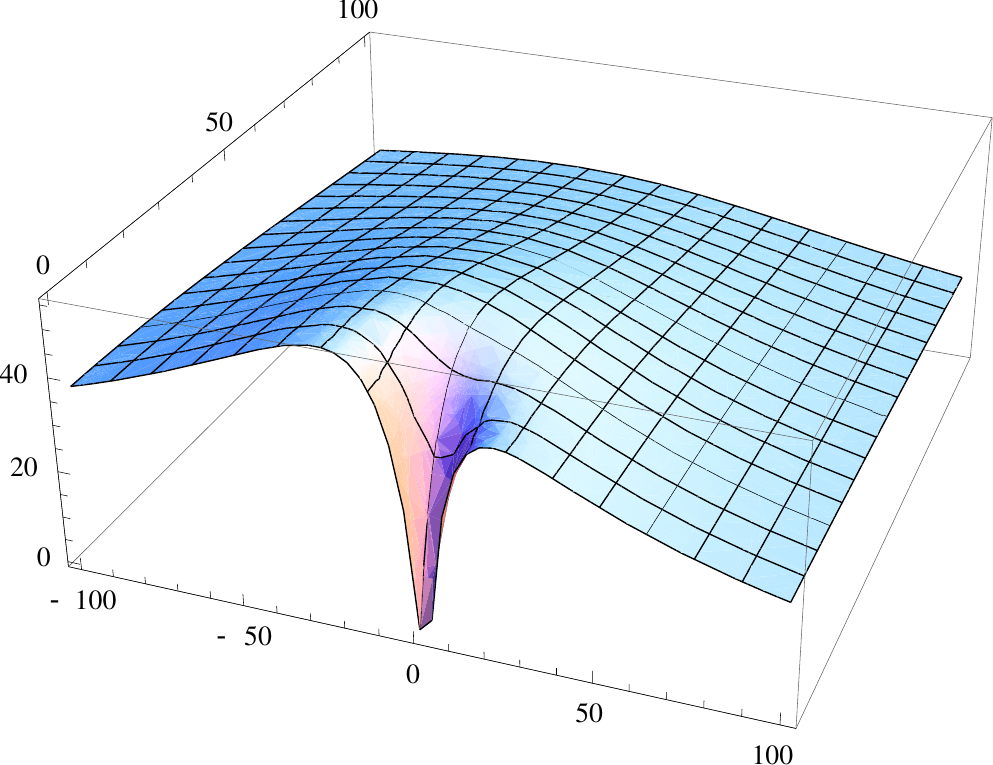}
\end{subfigure}
    \caption{\it \small The function $\cQ \equiv Z_1 Z_2 Z_3 w - w^2 \mu^2$ plotted near the source points at three different levels of magnification.  $\cQ$ is everywhere non-negative, and therefore the solution is free of CTC's.} 
\label{Q graphs}
\end{figure}

Finally, to show there are no CTC's anywhere, we plot 
\begin{equation}
\cQ \equiv Z_1 Z_2 Z_3 w - w^2 \mu^2
\end{equation}
in  \cref{Q graphs}.  We see that it is positive near the centers as we expect, and appears to be positive everywhere, giving us a supergravity solution which is globally free of closed timelike curves\footnote{Naturally, it is not enough just to look at graphs.  It is also helpful to plot $\cQ - \abs{\cQ}$, which quickly reveals any place $\cQ$ might go negative.  This was checked in this example, and $\cQ \geq 0$ everywhere.}.

\section{Conclusions}

Using the floating brane ansatz of \cite{Bena:2009fi} we have obtained a new, infinite family of solutions to 5-dimensional $\Neql 2$ ungauged supergravity coupled to two vector multiplets.  To build the solutions, we start with a LeBrun metric for the 4-dimensional base.  These metrics are K\"ahler and solve the Euclidean-Einstein-Maxwell equations, and are specified by two functions that solve the $SU(\infty)$ Toda equation and its linearization.  The full supergravity solution is then constructed by solving the ``floating brane equations" on this base space.  To these equations we obtain general, explicit solutions which generically represent a collection of concentric black rings stabilized by their angular momentum and electromagnetic charges.  Under appropriate regularity conditions, the black rings are replaced by topological ``bubbles", and the solutions are smooth and horizon-free.  Imposing causality conditions, we obtain ``bubble equations" which dictate the sizes of topological bubbles in terms of the cohomological fluxes trapped on them.

The 4-dimensional K\"ahler base space is interesting in its own right, and we spend some time analyzing its properties.  Choosing a subclass of LeBrun metrics with $U(1) \times U(1)$ symmetry, we are able to solve the Toda equation and write down an explicit metric.  Like the Gibbons-Hawking metrics, these metrics have an explicit $U(1)$ fiber that pinches off at various points along the axis to create a series of homological 2-spheres.  However, a new feature of the LeBrun metrics is that homological 2-spheres can also be formed by the other angular coordinate, and we obtain the specific boundary conditions that allow this to happen.  We also find a new feature as we approach the Gibbons-Hawking points, or ``geometric charges."  In the GH metric, the $U(1)$ near these points fibers over the $S^2$ in the base to give $S^3 / \ZZ_q$, which makes the local metric an orbifold $\RR^4 / \ZZ_q$.  In the LeBrun metric, however, one generically has $\RR^4 / G$ at these points, where $G \simeq \ZZ_m \times \ZZ_n$ acts on the two angular coordinates in $\RR^4 \simeq \RR^2 \times \RR^2$.  Finally, and perhaps most importantly, the explicit LeBrun metrics obtained have a Maxwell field which is non-trivially threaded through its various 2-cycles.  This allows rich new phenomena in the full supergravity solution that were not present in previous work by the author and collaborators \cite{Bobev:2011kk, Bobev:2012af}.

Looking at the full supergravity solution, we see a striking similarity between these non-supersymmetric solutions and the previous, well-known BPS solutions \cite{Bena:2007kg}.  The regularity conditions take virtually the same form.  By demanding the absence of CTC's, we also obtain ``bubble equations" which have largely the same features as in the BPS solutions:  a 2-cycle is held open by the product of the three flavors of fluxes threading it.  However, the non-BPS bubble equations at a given point involve not only the fluxes on cycles adjacent to that point, but also involve all the fluxes on the nonadjacent cycles (which is a radical departure from the BPS bubble equations).  This indicates \emph{new physics} that was not present in the BPS case, involving a richer variety of $E \times B$ type interactions.

It is known from previous work that these 5-dimensional non-supersymmetric solutions on a K\"ahler base are actually trivial KK reductions of \emph{BPS} solutions in the 6-dimensional IIB frame \cite{Gutowski:2003rg, Cariglia:2004kk, Bena:2011dd}.  This explains some of the features we see, and yet makes others more mysterious.  It seems clear that the 5-dimensional solutions are force-balanced by a kind of ``supersymmetry without supersymmetry" \cite{Duff:1997qz}, and in fact might be \emph{closer} to BPS than the so-called ``almost BPS" solutions \cite{Goldstein:2008fq, Bena:2009ev, Bena:2009en}.  For example, the bubble equations here and in the traditional 5d BPS solutions are both linear in the inverse distances $1/r_{ij}$, whereas the ``almost BPS" bubble equations are cubic in the inverse distances.  Still, there are important differences between these bubble equations and the 5d BPS bubble equations that must be explained if we are to think of these as ``secretly BPS."

Having found the non-BPS bubble equations, we also find that imposing the asymptotics of the near-horizon BMPV metric \cite{Breckenridge:1996is} precludes the existence of any finitely-sized bubbled solutions.  However, one can see the near-horizon BMPV-like metrics as the result of a limiting process of scaling solutions \cite{Bena:2007qc,Bena:2006kb,Vasilakis:2011ki}.  Alternatively, one can lift to the 6d IIB frame where one can allow different asymptotic behavior in one of the warp factors, and in this case one can find an infinite family of smooth geometries, with finitely-sized bubbles held open by their cohomological fluxes, which are asymptotic to a momentum wave solution on $AdS_3 \times S^3$.

It would be interesting to explore further the lift to the 6d IIB frame, as was done with the LeBrun-Burns metrics in \cite{Bobev:2012af}.  In 6 dimensions, one has the possibility of regular supertubes, and one might also get a better handle on why the bubble equations differ between here and the traditional setting (particularly in containing non-local interactions).

It would also be interesting to look for an asymptotically-flat completion of these solutions in 5 dimensions by relaxing the simplifying assumptions used in the floating brane ansatz \cite{Bena:2009fi}.  This is certainly a non-trivial thing to do, as one will likely be forced to address the full Einstein equations.

Finally, we also point out that while this work has focused on smooth solutions, one also has within the same solution set an infinite family of singular solutions, representing various collections of non-supersymmetric, yet force-balanced, spinning 3-charge black rings.

We have presented here a number of results and techniques which we hope yield insight into supergravity and black hole microstates.  Recent progress in the ability to find supergravity solutions is very exciting and full of possibilities, and it is clear that there are many avenues waiting to be explored.

\section*{Acknowledgements}

I would like to thank Nick Warner, Iosif Bena, Juan Maldacena, and Gary Horowitz for useful conversations.  I am also grateful to IPhT, CEA-Saclay, the Simons Foundation, the organizers of the String Theory and Gravity workshops in Benasque, Spain, and the organizers of the ``Fuzz or Fire?" workshop at KITP, Santa Barbara, for their hospitality and stimulating atmosphere where parts of this work were completed.  This work is supported in part by the DOE grant DE-FG03-84ER-40168.

\section*{Appendices}
\label{lebrun axi appendix}

\begin{appendices}

\section{Groups at conical points from lattices in $SO(4)$}
\label{orbifold groups}

In this section we discuss how to compute the orbifold structure at the conical singularities of the LeBrun metrics.  We stress that not every conical singularity is an orbifold singularity.  For a point to be an orbifold singularity, the geometry must approach $\RR^4 / G$ for some finite group $G \subset SO(4)$; however, for generic values of the parameters, one can also obtain more general conical singularities that cannot be locally modeled as a quotient space of $\RR^4$.  To illustrate the difference, consider two different 2-dimensional cone metrics:
\begin{equation} \label{cone metrics}
\dd s_A^2 = \dd r^2 + r^2 \frac{\dd \theta^2}{n^2}, \qquad \dd s_B^2 = \dd r^2 + r^2 \frac{m^2 \, \dd \theta^2}{n^2}, \qquad \theta \sim \theta + 2\pi,
\end{equation}
for $m, n > 0 \in \ZZ$ relatively prime\footnotemark{}.  In the first metric $\dd s_A^2$, a circuit around the tip of the cone subtends $2\pi/n$ radians; hence an $n$-fold cover of this space will fill out the standard $\RR^2$, and this is the quotient space $\RR^2 / \ZZ_n$.  In the second metric $\dd s_B^2$, however, a path enclosing the origin subtends $2\pi m / n$ radians, and there is no $p$-fold cover of this space that gives us $\RR^2$; hence it is not a quotient of $\RR^2$, and not, strictly speaking, an orbifold.  A similar phenomenon affects LeBrun metrics, except that there are two angular coordinates involved rather than one.

\subsection{Orbifold points and more general conical singularities}
\label{lattice method}

Near each conical point in the LeBrun metric, one finds that the (local) metric approaches that of flat $\RR^4$, but with the $U(1) \times U(1)$ coordinates identified on a lattice $\widetilde \Gamma$ different from the usual one  $\Gamma$.  One can define a group structure $G$, which is a finite subgroup of $U(1) \times U(1) \subset SO(4)$, by comparing the two lattices $\Gamma, \widetilde \Gamma$.  The conical point is an \emph{orbifold} point precisely when $\Gamma \subseteq \widetilde \Gamma$ as a sublattice, and then the local geometry approaches $\RR^4 / G$.  In this section we will compute $G$.

Let $\Gamma$ be the \emph{standard} lattice on which to identify the $U(1) \times U(1)$ coordinates of $\RR^4$.  In the coordinates
\begin{equation}
\dd s^2 (\RR^4) = \dd \rho^2 + \rho^2 \Big( \dd \theta^2 + \cos^2 \theta \, \dd \alpha^2 + \sin^2 \theta \, \dd \beta^2 \Big),
\end{equation}
one has $(\alpha, \beta) \sim (\alpha + 2\pi, \beta) \sim (\alpha, \beta+2\pi)$, and hence the basis $\Lambda$ of $\Gamma$ can be written
\begin{equation}
\Lambda = 2 \pi \begin{pmatrix} 1 & 0 \\ 0 & 1 \end{pmatrix},
\end{equation}
where the columns are the two basis vectors.  We note that $\Lambda$ is only defined up to right action by $GL(2,\ZZ)$, because we are free to choose any two column vectors that generate the same lattice.

We should then compare this lattice $\Gamma$ to the lattice $\widetilde \Gamma$ of coordinate identifications obtained from the near-singularity limit of the LeBrun metric (after transforming it into the same $\RR^4$ coordinates as above).

\subsubsection{Reduction to Smith normal form}

\begin{figure}
\centering
\includegraphics[height=4cm]{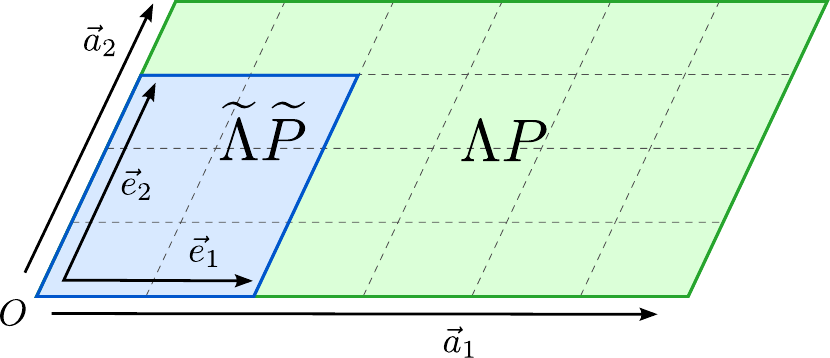}
    \caption{\it \small The lattice bases $\Lambda P$ and $\widetilde \Lambda \widetilde P$ are parallel.  There exist rational numbers $r_1, r_2$ such that $\vec a_1 = r_1 \, \vec e_1$ and $\vec a_2 = r_2 \, \vec e_2$.  In this case $r_1 = 3$ and $r_2 = 4/3$.} 
\label{lattices}
\end{figure}

The lattices $\Gamma, \widetilde \Gamma$ have unit cells which are parallelograms of any dimensions and oriented in any directions.  Let $\Lambda, \widetilde \Lambda$ be a choice of basis for each of $\Gamma, \widetilde \Gamma$.  Since the lattices are rational to each other, we can always make a change of basis via right action by $P, \widetilde P \in GL(2,\ZZ)$ such that the new bases $\Lambda P, \widetilde \Lambda \widetilde P$ are \emph{parallel}, by which we mean
\begin{equation} \label{parallel}
\widetilde \Lambda \widetilde P R = \Lambda P, \qquad \text{where} \qquad R = \begin{pmatrix} r_1 & 0 \\ 0 & r_2 \end{pmatrix},
\end{equation}
for some rational numbers $r_1, r_2 > 0$.  This is shown in \cref{lattices}.

The rational numbers $r_1, r_2$ give the factors by which each leg of $\Lambda P$ is larger than the same leg of $\widetilde \Lambda \widetilde P$.  It is easy to see that each leg of $\widetilde \Lambda \widetilde P$ generates a cyclic group modulo the unit cell $\Lambda P$, and hence one has
\begin{equation} \label{group G}
G \simeq \ZZ_m \times \ZZ_n, \quad \text{where} \quad m = \frac{r_1}{\gcd(1, r_1)}, \quad n = \frac{r_2}{\gcd(1, r_2)}.
\end{equation}

An \emph{orbifold} point occurs precisely when $r_1, r_2$ are integers, in which case the lattice cell $\widetilde \Lambda$ ``fits into'' $\Lambda$ evenly.  Then \cref{group G} can be written simply
\begin{equation} \label{orbi group G}
G \simeq \ZZ_m \times \ZZ_n, \quad \text{where} \quad m = r_1, \quad n = r_2.
\end{equation}
That is, at an orbifold point, the entries in the diagonal matrix $R$ give the orders of $\ZZ_m, \ZZ_n$.

What is left is to find $r_1, r_2$ in the first place.  To do this, one takes \cref{parallel} and isolates the diagonal matrix $R$:
\begin{equation} \label{basis comp}
R = \widetilde P^{-1} \widetilde \Lambda^{-1} \Lambda P.
\end{equation}
We do not need to know $P, \widetilde P \in GL(2,\ZZ)$ explicitly; we merely need to describe an algorithm for diagonalizing $\widetilde \Lambda^{-1} \Lambda$ by independent actions of $GL(2,\ZZ)$ from both the left and the right.  This is precisely the algorithm for finding the \emph{Smith normal form} of a matrix.  Since we have available both left and right $GL(2,\ZZ)$ actions, we may apply any sequence of elementary row \emph{or} column operations which are invertible over $\ZZ$.

Hence to obtain $R$ we diagonalize $\widetilde \Lambda^{-1} \Lambda$ via the following process.  At every step of the algorithm, we may
\begin{quote}
\begin{enumerate}
\item Swap any two rows or any two columns, or
\item Multiply any row, or any column, by $-1$, or
\item Add an integer multiple of any row (column) to another row (column).
\end{enumerate}
\end{quote}
The objective is to reach a diagonal matrix (this is always possible).  The full algorithm for the Smith normal form continues until the matrix is not only diagonal, but each entry along the diagonal divides the next, i.e. $r_1 | r_2$ in this case.  For our purposes, however, any diagonal matrix will do (and the result may not be unique).

In the case where the result is not unique, different possible results $R$ yield different ways of writing the \emph{same} group $G$.  For example, a given matrix might be diagonalized in two different ways to give $G \simeq \ZZ_4 \times \ZZ_6$ or $G \simeq \ZZ_2 \times \ZZ_{12}$, but these groups are isomorphic.  The same matrix cannot also be diagonalized to give, e.g. $\ZZ_3 \times \ZZ_8$---the algorithm as constructed preserves the group structure\footnotemark{}.

\footnotetext{Specifically, the reduction to Smith normal form of a square matrix $M$ preserves the sequence of \emph{invariant factors} $r_1 | r_2 | \ldots | r_n$ such that $\det M = r_1r_2\ldots r_n$ and each $r_i | r_{i+1}$.  It is precisely this sequence that distinguishes when the direct product of cyclic groups $\ZZ_{r_1} \times \ZZ_{r_2} \times \ldots \times \ZZ_{r_n}$ is isomorphic to another direct product of the same order.}

Once we have obtained $R$, we can then calculate the group $G$ via \cref{group G}.  We note that the order of $G$ is
\begin{equation}
\# G = mn = \frac{r_1}{\gcd(1, r_1)} \times \frac{r_2}{\gcd(1, r_2)} \geq \frac{r_1 r_2}{\gcd(1, r_1 r_2)}.
\end{equation}
But $r_1 r_2 = \det R = \det (\widetilde \Lambda^{-1} \Lambda)$.  Hence in terms of our lattice bases, we can put a lower bound on $\# G$:
\begin{equation} \label{G order lambda}
\# G \geq \frac{\det \Lambda}{\gcd ( \det \Lambda, \, \det \widetilde \Lambda )},
\end{equation}
where we assume, without loss of generality, that $\det \Lambda, \det \widetilde \Lambda > 0$ (which can always be arranged by the right action of $GL(2,\ZZ)$).  We note further that, at an \emph{orbifold} point where $r_1, r_2 \in \ZZ$, the inequality \cref{G order lambda} is saturated, and then we can calculate the order of the group $G$ directly from the invariants $\det \Lambda, \det \widetilde \Lambda$.

\subsection{The conical points of LeBrun metrics}
\label{orbi group results}

In this section we will find the groups $G$ at the conical points of the LeBrun metric using the methods outlined in the previous section.

Near the conical points, the LeBrun metric approaches the form \cref{source point metric}
\begin{equation}
\dd s^2(LB) = \dd \varrho^2 + \frac{\varrho^2}{4} \bigg[ \dd \theta^2 + \frac{1}{\widetilde q_\ell{}^2} \bigg( \widetilde{K}(\theta) \, \dd \tau^2 - 2 \widetilde{KQ}(\theta) \, \dd \tau \, \dd \phi + \widetilde{Q}(\theta) \, \dd \phi^2 \bigg) \bigg],
\label{LB orbi points}
\end{equation}
and one must then compare it to a standard flat metric on $\RR^4$,
\begin{equation}
\dd s^2 (\RR^4) = \dd \varrho^2 + \varrho^2 \Big( \dd \vartheta^2 + \cos^2 \vartheta \, \dd \alpha^2 + \sin^2 \vartheta \, \dd \beta^2 \Big),
\end{equation}
where $\theta = 2 \vartheta$.  From the LeBrun coordinates $(\tau, \phi)$, one can go to $(\alpha,\beta)$ via
\begin{align} 
\alpha &= \frac{1}{2 \widetilde q_\ell} \Big( (k^3_\ell + \bar K^3_\ell - k^3_0) \, \tau - (q_\ell + \bar Q_\ell - q_0) \, \phi \Big), \label{LB to R4 2a} \\
\beta &= \frac{1}{2 \widetilde q_\ell} \Big( (k^3_\ell - \bar K^3_\ell + k^3_0) \, \tau - (q_\ell - \bar Q_\ell + q_0) \, \phi \Big). \label{LB to R4 2b}
\end{align}

We need to define a ``standard'' lattice $\Gamma_{LB}$ on which the LeBrun coordinates $(\tau, \phi)$ should be identified in the first place.  This is actually an arbitrary choice (it will merely affect how we interpret the various parameters $q_\ell, k^3_\ell$).  But it is natural to borrow the standard ``diamond lattice'' from Gibbons-Hawking metrics:
\begin{equation} \label{tau phi ident}
(\tau, \phi) : \quad (0,0) \sim (4\pi,0) \sim (2\pi,2\pi) \sim (2\pi,-2\pi).
\end{equation}

By following the identifications \cref{tau phi ident} along the coordinate transformation \cref{LB to R4 2a}, \cref{LB to R4 2b}, we obtain the lattice $\widetilde \Gamma$ in the coordinates $(\alpha, \beta)$ given by the basis
\begin{equation} \label{lambdat}
\widetilde \Lambda = 2\pi \cdot \frac{1}{2 \widetilde q_\ell}
\begin{pmatrix}
k^3_\ell + \hat K^3_\ell + q_\ell + \hat Q_\ell \; & k^3_\ell + \hat K^3_\ell - q_\ell - \hat Q_\ell \\
k^3_\ell - \hat K^3_\ell + q_\ell - \hat Q_\ell \; & k^3_\ell - \hat K^3_\ell - q_\ell + \hat Q_\ell
\end{pmatrix},
\end{equation}
where for ease of legibility we have defined
\begin{equation}
\hat K^3_\ell \equiv \bar K^3_\ell - k^3_0, \qquad \hat Q_\ell \equiv \bar Q_\ell - q_0.
\end{equation}
The standard lattice $\Gamma$ in the coordinates $(\alpha, \beta)$ is given simply by the basis
\begin{equation}
\Lambda = 2\pi \begin{pmatrix} 1 & 0 \\ 0 & 1 \end{pmatrix},
\end{equation}
which makes the calculations easy, as $\widetilde \Lambda^{-1} \Lambda$ is just $2\pi \widetilde \Lambda^{-1}$.

From \cref{G order lambda}, we see that the order of the group $G$ is at least $\abs{\widetilde q_\ell}$:
\begin{equation}
\det(\widetilde \Lambda^{-1} \Lambda) = -\widetilde q_\ell, \quad \text{and hence} \quad \# G \geq \abs{\widetilde q_\ell},
\end{equation}
And if $r_1, r_2 \in \ZZ$, we have simply
\begin{equation}
\# G = \abs{\widetilde q_\ell} \quad \text{at orbifold points.}
\end{equation}

\subsubsection{When is a conical point an orbifold point?}

As we have pointed out, an orbifold point occurs when $r_1, r_2 \in \ZZ$, or alternatively, when $\widetilde \Lambda^{-1} \Lambda \in \mathrm{Mat}_2(\ZZ)$, the set (not group) of $2 \times 2$ matrices with integer entries.  This yields the condition
\begin{equation}
\frac{1}{2}
\begin{pmatrix}
k^3_\ell - \hat K^3_\ell - q_\ell + \hat Q_\ell \; & - k^3_\ell - \hat K^3_\ell + q_\ell + \hat Q_\ell \\
- k^3_\ell + \hat K^3_\ell - q_\ell + \hat Q_\ell \; & k^3_\ell + \hat K^3_\ell + q_\ell + \hat Q_\ell
\end{pmatrix}
\in \mathrm{Mat}_2(\ZZ),
\end{equation}
where notably the $1/\widetilde q_\ell$ in \cref{lambdat} has dropped out.  Thus a LeBrun metric contains \emph{only} orbifold points, and no generic conical points, when the sum of all the parameters is even:
\begin{equation} \label{parity cond app}
\bigg( k^3_0 + \sum_{i=1}^N k^3_i + q_0 + \sum_{i=1}^N q_i \bigg) \in 2\ZZ.
\end{equation}
Conversely, \emph{none} of the conical points have the quotient structure $\RR^4 / G$ if the sum of parameters is odd.  We will assume this sum is even such that each conical point is an orbifold point with structure $\RR^4 / G$.

\subsubsection{When is the group $G$ trivial?}

The group $G$ is trivial whenever $\widetilde \Gamma, \Gamma$ are the \emph{same} lattice.  This happens whenever $\widetilde \Lambda^{-1} \Lambda \in GL(2,\ZZ)$.  That is,
\begin{equation}
\frac{1}{2}
\begin{pmatrix}
k^3_\ell - \hat K^3_\ell - q_\ell + \hat Q_\ell \; & - k^3_\ell - \hat K^3_\ell + q_\ell + \hat Q_\ell \\
- k^3_\ell + \hat K^3_\ell - q_\ell + \hat Q_\ell \; & k^3_\ell + \hat K^3_\ell + q_\ell + \hat Q_\ell
\end{pmatrix}
\in GL(2,\ZZ),
\end{equation}
The factor of $1/2$ imposes the parity condition \eqref{parity cond app}.  Furthermore, the determinant of this matrix is $\widetilde q_\ell \equiv q_\ell \hat K^3_\ell - k^3_\ell \hat Q_\ell$.  Therefore for the metric to locally look like $\RR^4$ with no conical singularity requires
\begin{equation}
\widetilde q_\ell = \pm 1.
\end{equation}

\subsubsection{When is the group $G$ like a Gibbons-Hawking orbifold group?}

A 1-center Gibbons-Hawking metric with ``charge'' $m$, written
\begin{equation}
\dd s^2(GH) = \frac{r}{m} \Big( \dd \psi + m \cos \theta \, \dd \chi \Big)^2 + \frac{m}{r} \Big( \dd r^2 + r^2 \, \dd \theta^2 + r^2 \sin^2 \theta \, \dd \chi^2 \Big),
\end{equation}
is a metric on the orbifold $\RR^4 / \ZZ_m$, where $\ZZ_m$ acts in the \emph{diagonal} $U(1)$ of the maximal torus $U(1) \times U(1) \in SO(4)$.  In $(\alpha, \beta)$ coordinates, this corresponds to the lattice $\Gamma_{GH}$ with basis
\begin{equation}
\Lambda_{GH} = 2\pi \begin{pmatrix} 1 & \frac{p}{m} \\ 0 & \frac{p}{m} \end{pmatrix},
\end{equation}
where $p$ and $m$ are relativaly prime.  The LeBrun metric then has a ``diagonal'' orbifold point whenever $\widetilde \Lambda^{-1} \Lambda_{GH} \in GL(2,\ZZ)$, or equivalently, whenever $\Lambda_{GH}^{-1} \widetilde \Lambda \in GL(2,\ZZ)$, since the determinant is $\pm 1$ in any case.  This requires first that
\begin{equation}
\det (\Lambda_{GH}^{-1} \widetilde \Lambda) = - \frac{m}{p \widetilde q_\ell} = \pm 1, \quad \text{or} \quad m = \pm p \widetilde q_\ell.
\end{equation}
But since $p$ and $m$ are relatively prime, we must have $p = 1$ and $\widetilde q_\ell = m$.  Next, writing out $\Lambda_{GH}^{-1} \widetilde \Lambda$ we have
\begin{equation}
\frac{1}{2 \widetilde q_\ell}
\begin{pmatrix}
2 ( \hat K^3_\ell + \hat Q_\ell) \; & 2 ( \hat K^3_\ell - \hat Q_\ell ) \\
\widetilde q_\ell (k^3_\ell - \hat K^3_\ell + q_\ell - \hat Q_\ell) \; & \widetilde q_\ell (k^3_\ell - \hat K^3_\ell - q_\ell + \hat Q_\ell)
\end{pmatrix}
\in GL(2,\ZZ).
\end{equation}
So again, the sum of all the parameters must be even, and one gets a ``diagonal'' orbifold point wherever
\begin{equation}
\frac{2 (\bar K^3_\ell - k^3_0)}{\widetilde q_\ell} \in \ZZ \quad \text{and} \quad \frac{2(\bar Q_\ell - q_0)}{\widetilde q_\ell} \in \ZZ.
\end{equation}
One may also consider $\ZZ_m$ acting in the ``anti-diagonal'' $U(1)$, which in $(\alpha, \beta)$ coordinates corresponds to the lattice $\Gamma_{\overline{GH}}$ with basis
\begin{equation}
\Lambda_{\overline{GH}} = 2\pi \begin{pmatrix} 1 & -\frac{1}{m} \\ 0 & \frac{1}{m} \end{pmatrix}.
\end{equation}
One can similarly show that these points occur for $\widetilde q_\ell = m$ and
\begin{equation}
\frac{2 \, k^3_\ell}{\widetilde q_\ell} \in \ZZ \quad \text{and} \quad \frac{2 \, q_\ell}{\widetilde q_\ell} \in \ZZ.
\end{equation}

\section{Solutions to the Floating Brane system}
\label{solutions}

In this section we will solve the Floating Brane equations on the axisymmetric LeBrun base.

First, the $L_1, L_2$ equations \eqref{L1 L2 eqns} are simply the linearized Toda equation, which we have already solved to obtain $w$.  We define ``potentials" in the same way as in \eqref{w potential},
\begin{equation}
L_1 = \partial_z \hat L_1, \qquad L_2 = \partial_z \hat L_2,
\end{equation}
such that $\hat L_1, \hat L_2$ solve the cylindrically-symmetric Laplace equation:
\begin{gather}
\hat L_1 = \ell_1^0 \log \rho + \sum_i \ell_1^i \, G_i (\rho, \eta), \qquad
\hat L_2 = \ell_2^0 \log \rho + \sum_i \ell_2^i \, G_i (\rho, \eta), \\
G_i(\rho, \eta) = \log \frac{\eta - \eta_i + \sqrt{\rho^2 + (\eta - \eta_i)^2}}{\rho},
\end{gather}
where sums are understood to run from 1 to $N$.  Then $L_1, L_2$ can be written
\begin{align}
L_1 &=  \frac{1}{\rho (V_{\rho \eta}^2 + V_{\eta \eta}^2)} \big( V_{\eta \eta} \, \hat L_{1,\rho} - V_{\rho \eta} \, \hat L_{1,\eta} \big), \\
L_2 &=  \frac{1}{\rho (V_{\rho \eta}^2 + V_{\eta \eta}^2)} \big( V_{\eta \eta} \, \hat L_{2,\rho} - V_{\rho \eta} \, \hat L_{2,\eta} \big).
\end{align}

The $K^1, K^2, M$ equations \eqref{K1 eqn}, \eqref{K2 eqn}, \eqref{M eqn} are all similar to each other.  On the left-hand side is the cylindrically-symmetric Laplace operator on $\RR^3$, and on the right-hand side is a product of two functions that solve the linearized Toda equation.  Writing down the obvious homogeneous part, and then making an appropriate guess to match the source terms, the solutions are
\begin{align}
K^1 &= k^1_0 + \sum_i \frac{k^1_i}{\Sigma_i} + \frac{1}{V_{\rho \eta}^2 + V_{\eta \eta}^2} \bigg( V_{\eta \eta} \, \big( \hat w_\eta \hat L_{2,\eta} - \hat w_\rho \hat L_{2,\rho} \big) + V_{\rho \eta} \, \big( \hat w_\eta \hat L_{2,\rho} + \hat w_\rho \hat L_{2,\eta} \big) \bigg), \\
K^2 &= k^2_0 + \sum_i \frac{k^2_i}{\Sigma_i} + \frac{1}{V_{\rho \eta}^2 + V_{\eta \eta}^2} \bigg( V_{\eta \eta} \, \big( \hat w_\eta \hat L_{1,\eta} - \hat w_\rho \hat L_{1,\rho} \big) + V_{\rho \eta} \, \big( \hat w_\eta \hat L_{1,\rho} + \hat w_\rho \hat L_{1,\eta} \big) \bigg), \\
M &= m_0 + \sum_i \frac{m_i}{\Sigma_i} + \frac12 \, \frac{1}{V_{\rho \eta}^2 + V_{\eta \eta}^2} \bigg( V_{\eta \eta} \, \big( \hat L_{1,\eta} \hat L_{2,\eta} - \hat L_{1,\rho} \hat L_{2,\rho} \big) + V_{\rho \eta} \, \big( \hat L_{1,\eta} \hat L_{2,\rho} + \hat L_{1,\rho} \hat L_{2,\eta} \big) \bigg),
\end{align}
where $\Sigma_i \equiv \sqrt{\rho^2 + (\eta - \eta_i)^2}$.

The $L_3$ equation offers no shortcuts.  After a tedious exercise, one can show its solution is
\begin{equation}
\begin{split}
L_3 &= \ell_3^0 - \ell_3^z \, \rho V_\rho + \sum_i \frac{1}{\Sigma_i} \big( k^3_0 \ell_3^i + \ell_1^0 k^1_i + \ell_2^0 k^2_i + 2 q_0 m_i \big) \\
& \qquad + \sum_{\substack{ij \\ i \neq j}} \frac{1}{\eta_i - \eta_j} \frac{\Sigma_i}{\Sigma_j} \big( k^3_i \ell_3^j + \ell_1^i k^1_j + \ell_2^i k^2_j + 2 q_i m_j \big) \\
& \qquad - \sum_i \frac{\eta - \eta_i}{\Sigma_i} \big( k^3_i \ell_3^i + \ell_1^i k^1_i + \ell_2^i k^2_i + 2 q_i m_i \big) \\
& \qquad + \frac{\rho}{V_{\rho \eta}^2 + V_{\eta \eta}^2} \bigg[ V_{\rho\eta} \, \Big( \! - \hat w_\eta \hat L_{1,\eta} \hat L_{2,\eta} + \hat w_\rho \hat L_{1,\rho} \hat L_{2,\eta} + \hat w_\rho \hat L_{1,\eta} \hat L_{2,\rho} + \hat w_\eta \hat L_{1,\rho} \hat L_{2,\rho} \Big) \\
& \qquad \qquad \qquad + V_{\eta\eta} \, \Big( \! - \hat w_\rho \hat L_{1,\rho} \hat L_{2,\rho} + \hat w_\rho \hat L_{1,\eta} \hat L_{2,\eta} + \hat w_\eta \hat L_{1,\rho} \hat L_{2,\eta} + \hat w_\eta \hat L_{1,\eta} \hat L_{2,\rho} \Big) \bigg],
\end{split}
\end{equation}
where the parameter $\ell_3^z$ multiplies $z = - \rho V_\rho$.  It is important to note here that the pair $k^3_i, \ell_3^j$ behaves oppositely to the pairs $\ell_1^i, k^1_j$ and $\ell_2^i, k^2_j$.

Finally, one must solve the $\omega$ equation \eqref{omega eqn}.  If we write
\begin{equation}
\omega = \omega_{(\phi)} \, \dd \phi,
\end{equation}
then \eqref{omega eqn} reduces to the two equations
\begin{align}
\begin{split}
r \partial_r \big( \omega_{(\phi)} \big) &= \frac12 \big( \rho^2 L_1 \, \partial_z K^1 - K^1 \, \partial_z (\rho^2 L_1) \big) + \frac12 \big( \rho^2 L_2 \, \partial_z K^2 - K^2 \, \partial_z (\rho^2 L_2) \big) \\
& \qquad + \frac14 \big( L_3 \, \partial_z^2 (\rho^2) - \partial_z (\rho^2) \, \partial_z L_3 \big) + \rho^2 w \, \partial_z M - M \, \partial_z (\rho^2 w) - 2 \rho^2 w L_1 L_2,
\end{split} \\
\begin{split}
- \partial_z \big( \omega_{(\phi)} \big) &= \frac12 \big( L_1 \, r \partial_r K^1 - K^1 \, r \partial_r L_1 \big) + \frac12 \big( L_2 \, r \partial_r K^2 - K^2 \, r \partial_r L_2 \big) \\
& \qquad + \frac14 \big( L_3 \, r \partial_r u_z - u_z \, r \partial_r L_3 \big) + w \, r \partial_r M - M \, r \partial_r w.
\end{split}
\end{align}
It is again a tedious exercise to show that these are solved by
\begin{align}
\omega_{(\phi)} &= \omega_0 + \frac{1}{\rho^2 (V_{\rho \eta}^2 + V_{\eta \eta}^2)} \bigg\{ \frac12 \ell_3^z \bigg( \rho^2 V_\rho V_{\rho \eta} - \eta \rho^2 (V_{\rho \eta}^2 + V_{\eta \eta}^2) \bigg) \\
& \qquad + \frac12 \big( k^1_0 \ell_1^0 + k^2_0 \ell_1^0 - \ell_3^0 + 2 m_0 q_0 \big) \bigg( k^3_0 - \sum_i \frac{\eta - \eta_i}{\Sigma_i} k^3_i \bigg) \notag \\
& \qquad - \frac12 k^3_0 \sum_i \big( k^1_0 \ell_1^i + k^2_0 \ell_2^i + 2 m_0 q_i \big) \frac{\eta - \eta_i}{\Sigma_i} \notag \\
& \qquad + \frac12 \sum_{ij} k^3_i \big( k^1_0 \ell_1^j + k^2_0 \ell_2^j + 2 m_0 q_j \big) \frac{\rho^2 + (\eta - \eta_i)(\eta - \eta_j)}{\Sigma_i \Sigma_j} \notag \\
& \qquad + \frac12 k^3_0 \sum_{\substack{ij \\ i \neq j}} \big( k^1_i \ell_1^j + k^2_i \ell_2^j - \ell_3^i k^3_j + 2 m_i q_j \big) \frac{1}{\eta_i - \eta_j} \frac{\rho^2 + (\eta - \eta_i)(\eta - \eta_j)}{\Sigma_i \Sigma_j} \notag \\
& \qquad -  \frac12 \sum_{\substack{ijk \\ i \neq j}} k^3_k \big( k^1_i \ell_1^j + k^2_i \ell_2^j + 2 m_i q_j \big) \frac{1}{\eta_i - \eta_j} \frac{1}{\Sigma_i \Sigma_j \Sigma_k} \times \notag \\
& \qquad \qquad \qquad \qquad \qquad \times \Big[ \rho^2 \big( \eta - \eta_i + \eta_j - \eta_k \big) + (\eta - \eta_i) (\eta - \eta_j) (\eta - \eta_k) \Big] \notag \\
& \qquad + \frac12 \sum_{ik} k^3_k \big( k^1_i \ell_1^i + k^2_i \ell_2^i + 2 m_i q_i \big) \frac{\rho^2}{\Sigma_i^2 \Sigma_k} \notag \\
& \qquad + \frac12 \sum_{\substack{ijk \\ i \neq k}} k^3_i k^3_j \ell_3^k \frac{\eta_i - \eta_j}{\eta_i - \eta_k} \frac{\rho^2}{\Sigma_i \Sigma_j \Sigma_k} - \frac12 \sum_{ij} k^3_i k^3_j \ell_3^i \frac{\rho^2}{\Sigma_i^2 \Sigma_j} + \frac12 \sum_i (k^3_i)^2 \ell_3^i \frac{\rho^2}{\Sigma_i^3} \notag \\
& \qquad + \frac12 \sum_{\substack{ijk \\ i \neq k}} k^3_i k^3_j \ell_3^k \frac{1}{\eta_i - \eta_k} \frac{(\eta - \eta_k) \big( \rho^2 + (\eta - \eta_i)(\eta - \eta_j) \big)}{\Sigma_i \Sigma_j \Sigma_k} \notag \\
& \qquad + \sum_{ijk} q_i \ell_1^j \ell_2^j \frac{\rho^2}{\Sigma_i \Sigma_j \Sigma_k} \bigg\}, \notag
\end{align}
where again, all sums are assumed to run over $i,j,k \in \{1 \ldots N\}$.

We now have the complete data for constructing supergravity solutions.  The solution is characterized by $N$ number of points $\eta_i$ along the axis in the base space, and by the $8N + 10$ parameters $\{q_0, k^1_0, k^2_0, k^3_0, \ell_1^0, \ell_2^0, \ell_3^0, m_0, \omega_0, \ell_3^z, q_i, k^1_i, k^2_i, k^3_i, \ell_1^i, \ell_2^i, \ell_3^i, m_i\}$, which in general are constrained by the requirement for the absence of CTC's and Dirac-Misner strings.  Finally, to complete the supergravity solution, one puts the functions $w, K^1, K^2, K^3, L_1, L_2, L_3, M$ into the ans\"atze of Sections \ref{floating branes} and \ref{lebrun floating branes}.

\subsection{Asymptotic expansions of the metric functions}
\label{asymptotics app}

In this section are the detailed asymptotic expansions of the metric functions in terms of the above solutions.

First, the parameters $k^1_0, k^2_0, \ell_3^z$ lead to terms that blow up at infinity, so we set 
\begin{equation} \label{no const}
k^1_0 = 0, \qquad k^2_0 = 0, \qquad \ell_3^z = 0.
\end{equation}
To look near infinity it is helpful to define the coordinates $R, \theta$ via
\begin{equation}
\rho = R \sin \theta, \qquad \eta = R \cos \theta.
\end{equation}
Then the warp factors $Z_1, Z_2$ go as $1/R$:
\begin{equation} \label{Z1 Z2 asym}
Z_1 \sim \bigg( \frac{K^2_\star K^3_\star + Q_\star L_1^\star}{q_0 K^3_\star - k^3_0 Q_\star} \bigg) \frac{1}{R}, \qquad Z_2 \sim \bigg( \frac{K^1_\star K^3_\star + Q_\star L_2^\star}{q_0 K^3_\star - k^3_0 Q_\star} \bigg) \frac{1}{R},
\end{equation}
where we define the quantities
\begin{gather}
K^1_\star \equiv \sum_{i=1}^N k^1_i, \qquad K^2_\star \equiv \sum_{i=1}^N k^2_i, \qquad K^3_\star \equiv \sum_{i=1}^N k^3_i, \qquad Q_\star \equiv \sum_{i=1}^N q_i, \\
L_1^\star \equiv \sum_{i=1}^N \ell_1^i, \qquad L_2^\star \equiv \sum_{i=1}^N \ell_2^i, \qquad L_3^\star \equiv \sum_{i=1}^N \ell_3^i, \qquad M_\star \equiv \sum_{i=1}^N m_i.
\end{gather}
At leading order, the remaining metric functions $Z_3, \mu, \omega_{(\phi)}$ go as constants:
\begin{align}
Z_3 &\sim \ell_3^0 - \sum_{\substack{ij \\ i \neq j}} \frac{k^1_i \ell_1^j + k^2_i \ell_2^j - k^3_i \ell_3^j + 2 m_i q_j}{\eta_i - \eta_j}, \label{Z3 1st} \\
\mu &\sim m_0 - \frac12 \, \frac{K^3_\star}{q_0 K^3_\star - k^3_0 Q_\star} \bigg( \ell_3^0 - \sum_{\substack{ij \\ i \neq j}} \frac{k^1_i \ell_1^j + k^2_i \ell_2^j - k^3_i \ell_3^j + 2 m_i q_j}{\eta_i - \eta_j} \bigg), \\
\begin{split}
\omega_{(\phi)} &\sim \omega_0 + \frac12 \, \frac{Q_\star}{q_0 K^3_\star - k^3_0 Q_\star} \bigg( \ell_3^0 - \sum_{\substack{ij \\ i \neq j}} \frac{k^1_i \ell_1^j + k^2_i \ell_2^j - k^3_i \ell_3^j + 2 m_i q_j}{\eta_i - \eta_j} \bigg) \\
& \qquad \qquad + \bigg( \frac{k^3_0 q_0 + K^3_\star Q_\star - (q_0 K^3_\star + k^3_0 Q_\star) \cos \theta}{(k^3_0)^2 + (K^3_\star)^2 - 2 \, k^3_0 K^3_\star \cos \theta} \bigg) \times \\
& \qquad \qquad \qquad \times \bigg[ m_0 - \frac12 \, \frac{K^3_\star}{q_0 K^3_\star - k^3_0 Q_\star} \bigg( \ell_3^0 - \sum_{\substack{ij \\ i \neq j}} \frac{k^1_i \ell_1^j + k^2_i \ell_2^j - k^3_i \ell_3^j + 2 m_i q_j}{\eta_i - \eta_j} \bigg) \bigg].
\end{split}
\end{align}
However, we must have $\mu \to 0, \; \omega_{(\phi)} \to 0$ asymptotically in order to avoid CTC's at infinity.  Therefore we must set
\begin{align}
m_0 &= \frac12 \frac{K^3_\star}{q_0 K^3_\star - k^3_0 Q_\star} \bigg( \ell_3^0 - \sum_{\substack{ij \\ i \neq j}} \frac{k^1_i \ell_1^j + k^2_i \ell_2^j - k^3_i \ell_3^j + 2 m_i q_j}{\eta_i - \eta_j} \bigg), \label{m0 soln} \\
\omega_0 &= - \frac12 \frac{Q_\star}{q_0 K^3_\star - k^3_0 Q_\star} \bigg( \ell_3^0 - \sum_{\substack{ij \\ i \neq j}} \frac{k^1_i \ell_1^j + k^2_i \ell_2^j - k^3_i \ell_3^j + 2 m_i q_j}{\eta_i - \eta_j} \bigg). \label{w0 soln}
\end{align}
Then in fact the asymptotic expansions of $\mu, \omega_{(\phi)}$ must be carried to the next order, giving
\begin{equation}
\begin{split}
\mu &\sim \frac{1}{R} \bigg\{  \frac{1}{(q_0 K^3_\star - k^3_0 Q_\star)^2} \bigg[ - K^3_\star \Big( (k^3_0)^2 + (K^3_\star)^2 - 2 \, k^3_0 K^3_\star \cos \theta \Big) K^1_\star K^2_\star \\
& \qquad \qquad \qquad - K^3_\star \Big( k^3_0 q_0 + K^3_\star Q_\star - 2 \, k^3_0 Q_\star \cos \theta \Big) \big( K^1_\star L_1^\star + K^2_\star L_2^\star \big) \\
& \qquad \qquad \qquad - Q_\star \Big( k^3_0 q_0 + K^3_\star Q_\star - ( q_0 K^3_\star + k^3_0 Q_\star) \cos \theta \Big) L^1_\star L^2_\star \bigg] \\
& \qquad \qquad + \frac12 \, \frac{1}{q_0 K^3_\star - k^3_0 Q_\star} \bigg[ \big( k^3_0 + K^3_\star \cos \theta \big) \big( K^1_\star L_1^\star + K^2_\star L_2^\star \big) \\
& \qquad \qquad \qquad \qquad + \big( K^3_\star \cos \theta - k^3_0 \big) \big( K^3_\star L_3^\star + 2 Q_\star M_\star \big) \bigg] \bigg\},
\end{split}
\end{equation}
and
\begin{equation}
\omega_{(\phi)} \sim \frac{1}{2 R} \, \frac{K^3_\star \sin^2 \theta}{(k^3_0)^2 + (K^3_\star)^2 - 2 \, k^3_0 K^3_\star \cos \theta} \big( K^1_\star L_1^\star + K^2_\star L_2^\star + K^3_\star L_3^\star + 2 Q_\star M_\star \big).
\end{equation}

\end{appendices}



\providecommand{\href}[2]{#2}\begingroup\raggedright\endgroup


\end{document}